\documentclass[aps,prd,reprint,groupedaddress, onecolumn]{revtex4-1}

\usepackage[utf8]{inputenc}
\usepackage[english]{babel}
\usepackage[T1]{fontenc}
\usepackage{amsmath}
\usepackage{amsfonts}
\usepackage{hyperref}
\usepackage{amssymb}
\usepackage{graphicx}
\usepackage{tikz-feynman}
\usepackage{subfig}
\usepackage{physics}
\usepackage{array}

\newcommand{\mpl}{M_\mathrm{P}}
\newcommand{\sch}{Schwarzschild }

\begin{document}


\title{Half-solution to the two-body problem in General Relativity}


\author{Adrien Kuntz}
\email[]{kuntz@cpt.univ-mrs.fr}
\affiliation{Aix Marseille Univ, Universit\'{e} de Toulon, CNRS, CPT, Marseille, France}


\date{\today}

\begin{abstract}
We show that the introduction of two worldline parameters defines a different approach to computations in the effective field theory approach to the two-body problem in General Relativity and present some preliminary evidence for a reduction in computational complexity. These parameters obey a polynomial equation whose perturbative expansion recovers an infinite series of diagrams. Futhermore, we show that our equations define an effective two-body horizon for interacting particles in General Relativity; in the circular orbit case, it corresponds to the smallest conceivable separation up to which the orbit can remain circular. We expect our results to simplify higher-order computations in the two-body problem, as well as to give insights on the nonperturbative properties of interacting binaries.

\end{abstract}


\maketitle


\section{Introduction}


Gravitational waves (GW) from compact binaries \cite{Abbott_2017,Abbott:2016blz} provide exciting new challenges for our understanding of gravity in the strong-field regime. Waveform modeling require precise theoretical predictions on the two-body dynamics in General Relativity (GR) \cite{Lindblom_2008} for the LIGO/Virgo data analysis \cite{2015, Acernese_2014}, as well as for the future interferometers such as LISA \cite{amaroseoane2017laser} and Einstein Telescope \cite{Sathyaprakash_2012}. The inspiral part is well described by the slow-motion, weak-field or post-Newtonian (PN) approximation \cite{Blanchet:2013haa}, while for the last stages of the dynamics Numerical Relativity is needed \cite{2009LRR....12....1G}. More recently, Effective Field Theory (EFT) ideas from particle physics have been applied to the two-body problem \cite{goldberger_effective_2006,porto_effective_2016,Levi:2018nxp} in the framework of Non-Relativistic General Relativity (NRGR). The perturbative computations of the post-Newtonian approximation have been successfully translated in a series of Feynman diagrams. The current state-of-the art computation in NRGR is at the 4PN level (i.e, $v^8$ beyond the Newtonian order where $v \ll 1$ is the typical velocity of the two bodies) \cite{Foffa_2013, Foffa_2017, 2019PhRvD.100b4047F,2019PhRvD.100b4048F}; this computation confirmed earlier results obtained in the PN formalism \cite{Damour_2014, Damour_2015, Damour_2016, Bernard_2016, Bernard_2017, Bernard_2017_2, Marchand_2018, Bernard_2018}. Furthermore, NRGR has been generalized to include spin and this yielded results which up to date have not been obtained by any other methods~\cite{Porto_2006, Levi:2015msa, Levi:2020kvb, Levi:2014gsa, Levi:2015ixa, Levi:2016ofk}.

In a parallel development, the field of scattering amplitudes \cite{2017arXiv170803872C} has come into contact with the post-Minkowskian (PM) approach to the two-body problem. Compared to the PN formalism, the PM philosophy is to keep the velocity arbitrary while still expanding observables for weak fields. This procedure is well adapted to scattering processes in which velocities may be high but such that one still restricts to large values of the impact parameter. The scattering angle is known at the 3PM order \cite{PhysRevLett.122.201603} and can be translated in conservative Hamiltonians for binary systems \cite{Cheung:2018wkq} allowing for a useful cross-check with the PN formalism within their overlapping domain of validity. Even if the PM computations are far from reaching the same perturbative order as the PN ones, one can argue that they provide a somewhat \textit{more exact} result at the same order since they are exact to all orders in velocity \cite{Antonelli:2019ytb}. In this article, we will adopt the PM philosophy in the sense that our results will be nonperturbative in some part of the dynamics, while still being perturbative in the other part. This is achieved by a simplification of the Feynman rules of NRGR, such that no perturbative assumption is made on the matter sector. Let us now be more precise about this statement.

In NRGR as well as in the PN formalism, the binary constituents (considered nonspinning here for simplicity) are modeled as point-particles. The Feynman diagrams of the two point-particles involve graviton vertices connected by propagators. The vertices are of two types: the \textit{bulk} nonlinearities originating from the Einstein-Hilbert action, and the \textit{worldline} couplings coming from the matter action describing the two point-particles. In this article, we point out that the introduction of two worldline parameters - or einbeins - allows to drastically simplify the worldline couplings, leaving an action containing only a \textit{linear} coupling of the graviton to the source. The number of Feynman diagrams needed to evaluate the action at each perturbative order is thus greatly reduced, so that the only computational obstacle in the two-body problem is entirely contained in the Einstein-Hilbert action. Such a procedure is the perfect analogue of going from the Nambu-Goto to the Polyakov action in string theory \cite{Polyakov}. Concerning the two-body problem, similar worldline parameters were also introduced in~\cite{Galley_2013} in the ultra-relativistic limit of NRGR  and in~\cite{2019arXiv191008831D} to resum a series of Feynman diagrams when considering disformally coupled scalar fields.

Next, we investigate in more details the physical properties of the worldline parameters which we introduced. In the case of circular orbits, these are known as the redshift variables \cite{Detweiler:2008ft} as they represent the redshift of a photon emitted close to the point-particles and detected at large distance from the system. We find that these variables obey a fifth-order polynomial equation whose properties are examined in both the static and circular orbit case. We show that this equation does not admit solutions for close enough binaries, so that it allows to define an 'effective two-body horizon' ; more precisely, we find that for a critical separation no circular orbit can exist at all for the two-body problem. This is a two-body generalization of the well-known Innermost Circular Orbit (ICO) \footnote{Defined as the smallest possible circular orbit ; not to be confused with the Innermost \textit{stable} Circular Orbit or ISCO } of the \sch geometry. This result could shed light on nonperturbatives properties of the two-body motion.

In the following we use units in which $\hbar = c = 1$, we define Planck's mass by $\mpl^2 = 1/(8 \pi G)$ where $G$ is Newton's constant, and we use the mostly plus metric convention.

\section{Integrating out gravity}

Let us begin by summarizing the NRGR approach to the two-body dynamics in GR \cite{goldberger_effective_2006,porto_effective_2016}. Along the way, we will introduce the worldline parameters. We take our action to be the one of GR, i.e
\begin{equation}
S = \frac{\mpl^2}{2} \int \mathrm{d}^4 x \sqrt{-g} R + S_{m,1} + S_{m,2} \; ,
\end{equation}
where the matter action is constituted of two point-particles $\alpha=1,2$,
\begin{equation} \label{eq:pp_action}
S_{m,\alpha} = - m_\alpha \int \mathrm{d}t \sqrt{-g_{\mu \nu} v_\alpha^\mu v_\alpha^\nu} \; , \quad v_\alpha^\mu = \frac{\mathrm{d} x_\alpha^\mu}{\mathrm{d} t} \;  .
\end{equation}
 In the NRGR formalism, the key quantity is the effective action formally defined as the path integral of the action,
\begin{equation}
e^{i S_\mathrm{eff}} = \int \mathcal{D} h_{\mu \nu} e^{iS} \; .
\end{equation}
The path integral is calculated with a Feynman expansion, by expanding the Ricci scalar and the point-particle action in powers of $h_{\mu \nu}$. The quadratic term defines the propagator of the gravitational field, while the infinite series due to the nonlinearities of GR give the vertices in the diagrams. The essential point of this article is that the nonlinearities associated to the point-particle action (we will refer to them as \textit{worldline} nonlinearities) can be computed exactly, by introducing two auxiliary parameters (we will give their physical interpretation in Section \ref{sec:circular}; notice that such parameters were also introduced in the ultra-relativistic limit of NRGR in \cite{Galley_2013}.). Let us rewrite each point-particle action as
\begin{equation} \label{eq:pp_action_e}
S_{m,\alpha} = - \frac{m_\alpha}{2} \int \mathrm{d}t \left[ e_\alpha - \frac{g_{\mu \nu} v_\alpha^\mu v_\alpha^\nu}{e_\alpha} \right] \; .
\end{equation}
Variation with respect to the einbein gives $e_\alpha = \sqrt{-g_{\mu \nu} v_\alpha^\mu v_\alpha^\nu}$
which yields back the original point-particle action \ref{eq:pp_action}. Instead, we will keep the $e_\alpha$ undetermined from now on and integrate out the gravitational field. The crucial improvement that this procedure yields is that the point-particle vertex is now \textit{linear} in the gravitational field, allowing for an \textit{exact} computation of the effective action.

Let us split $g_{\mu \nu} = \eta_{\mu \nu} + h_{\mu \nu}/ \mpl$. We add to the action the following harmonic gauge-fixing term \cite{goldberger_effective_2006} (so that our results will be expressed in harmonic coordinates):
\begin{equation}
S_\mathrm{GF} = - \frac{\mpl^2}{4} \int \mathrm{d}^4 x \sqrt{-g} \Gamma_\mu \Gamma^\mu \; , \quad \Gamma^\mu = \Gamma^\mu_{\alpha \beta} g^{\alpha \beta} \; ,
\end{equation}
$\Gamma^\mu_{\alpha \beta}$ being the Christoffel symbols for the metric $g_{\mu \nu}$.
Expanding in $ h_{\mu \nu}/ \mpl$, the total quadratic term in the action is
\begin{equation} \label{eq:EH_second_order}
S^{(2)} = - \frac{1}{8} \int \mathrm{d}^4x \left[ -\frac{1}{2}(\partial_\mu h^\alpha_\alpha)^2 + (\partial_\mu h_{\nu \rho})^2 \right] \;,
\end{equation}
which defines the propagator,
\begin{equation}
\left\langle T h_{\mu \nu}(x) h_{\alpha \beta}(x') \right\rangle = D_F(x-x') P_{\mu \nu ; \alpha \beta} \;,
\label{eq:propagator}
\end{equation}
where $T$ denotes time ordering, the Feynman propagator $D_F(x-x')$ is given by 
\begin{equation}
D_F(x_1-x_2) = \int \frac{d^4 k}{(2 \pi)^4} \frac{-i}{k^2-i\epsilon} e^{-ik(x_1-x_2)} \; ,
\end{equation}
the term  $i\epsilon$ is the prescription for the contour integral, and the tensor $P_{\mu \nu ; \alpha \beta}$ is
\begin{equation}
P_{\mu \nu ; \alpha \beta} = 2 \left( \eta_{\mu \alpha} \eta_{\nu \beta} + \eta_{\mu \beta} \eta_{\nu \alpha} - \eta_{\mu \nu} \eta_{\alpha \beta} \right) \; .
\end{equation}
The real part of $i D_F$, which contains all the information needed to extract the conservative dynamics of the system, can be computed using $1/(k^2-i\epsilon) = PV(1/k^2) + i \pi \delta(k^2)$ and reads
\begin{equation}  \label{eq:Feyprop}
\Re i D_F(x_1-x_2) = \frac{1}{8 \pi \vert \mathbf{x}_1 - \mathbf{x}_2 \vert} \left( \delta\left(t_1 - t_2 - \vert \mathbf{x}_1 - \mathbf{x}_2 \vert \right) +  \delta\left(t_1 - t_2 + \vert \mathbf{x}_1 - \mathbf{x}_2 \vert \right) \right) \; .
\end{equation}

The theory defined by eqs. \eqref{eq:EH_second_order} and \eqref{eq:pp_action_e} is now a simple quadratic theory linearly coupled to two sources. We can thus obtain exactly the effective action. Of course this neglects the higher-order vertices in $h_{\mu \nu}$ arising from the expansion of the Einstein-Hilbert action and from the gauge-fixing term (we call them \textit{bulk} nonlinearities in the following). These vertices are suppressed by 1PN order in the PN expansion, so that strictly speaking our computation will be of 0PN order. On the other hand our result should generalize the 1PM results \cite{Cheung:2018wkq}, since the 1PM order consists in taking the full graviton propagator \eqref{eq:Feyprop} with a linearized source term \cite{Foffa:2013gja}.

\begin{figure}
	\centering
%
%
\includegraphics[scale=1]{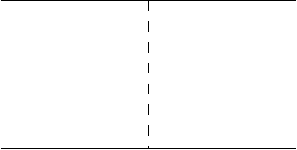}

\caption{The only Feynman diagram arising from the linear coupling \ref{eq:matter_coupling_Lagrange}. The dotted line represents an insertion of the propagator \ref{eq:Feyprop}, while the continuous lines represent point-particles treated as external sources }
\label{fig:one_feynman}
\end{figure}

  We rewrite the point-particle action as
\begin{align}
S_{m,\alpha} &= - \frac{m_\alpha}{2} \int \mathrm{d}t \left[e_\alpha + \frac{1-v_\alpha^2}{e_\alpha} \right] \\
&+ \frac{m_\alpha}{2 \mpl} \int \frac{\mathrm{d}t}{e_\alpha(t)} h_{\mu \nu} v_\alpha^\mu v_\alpha^\nu \; , \label{eq:matter_coupling_Lagrange}
\end{align}
where the first line does not depend on the gravitational field. The couplings of the graviton to the point-particle give rise to \textit{only} one Feynman diagram, represented in Figure \ref{fig:one_feynman}. It is easily computed as
\begin{equation}
\left. i S_\mathrm{eff} \right\vert _{\ref{fig:one_feynman}} = \frac{i m_1}{2 \mpl} \frac{i m_2}{2 \mpl} \int \frac{\mathrm{d} t_1}{e_1(t_1)} \frac{\mathrm{d} t_2}{e_2(t_2)} P_{\mu \nu; \alpha \beta} v_1^\mu v_1^\nu v_2^\alpha v_2^\beta D_F \left( t_1-t_2, \mathbf{x}_1(t_1) - \mathbf{x}_2(t_2) \right) \; .
\end{equation}

We will ignore retardation effects as a first step, and show after how to include them. This means that we set $t=t'$ in the Feynman propagator \ref{eq:Feyprop}, giving the final effective action where the gravitational field has been effectively removed
\begin{equation} \label{eq:my_Lagrangian}
S_\mathrm{eff} = \int \mathrm{d} t \left[ - \frac{m_1}{2} \left(e_1 + \frac{1-v_1^2}{e_1} \right) - \frac{m_2}{2} \left(e_2 + \frac{1-v_2^2}{e_2} \right) + \frac{\lambda G m_1 m_2 }{e_1 e_2 r} \right] \; ,
\end{equation}
where $r = \vert  \mathbf{x}_1 - \mathbf{x}_2 \vert$ and $\lambda$ is a combination of the two velocities,
\begin{equation}
\lambda = 1 + v_1^2 + v_2^2 - 4 \mathbf{v}_1 \cdot \mathbf{v}_2 - v_1^2 v_2^2 + 2 (\mathbf{v}_1 \cdot \mathbf{v}_2)^2 \; . 
\end{equation}

  In the following, we will sometimes refer to the first two terms of this equation as the "kinetic" Lagrangian, and the last one as the "interaction" Lagrangian, even if the separation between kinetic and potential energy is only valid in the Newtonian limit.

It is straightforward to generalize the precedent result to include retardation effects. The first delta function in the Feynman propagator fixes the time $t_2$ to be the retarded time, defined by the equation
\begin{equation}
t_2^R = t_1 - \left \vert \mathbf{x}_1(t_1) - \mathbf{x}_2(t_2^R) \right \vert \; ,
\end{equation}
and the second delta function is just a relabeling $1 \leftrightarrow 2$. Taking into account the Jacobian in the delta function, we obtain the action
\begin{align}
\begin{split} \label{eq:my_eq_retarded}
S_\mathrm{eff} &= \int \mathrm{d} t \left[ - \frac{m_1}{2} \left(e_1(t) + \frac{1-v_1^2}{e_1(t)} \right) - \frac{m_2}{2} \left(e_2(t) + \frac{1-v_2^2}{e_2(t)} \right) \right. \\
& + \left. \frac{\lambda(t, t^R) G m_1 m_2 }{2 e_1(t) e_2(t^R) \left[ \left \vert \mathbf{x}_1(t) - \mathbf{x}_2(t^R) \right \vert - \mathbf{v}_2(t^R) \cdot (\mathbf{x}_1(t) - \mathbf{x}_2(t^R))\right]} + (1 \leftrightarrow 2) \right] \; ,
\end{split}
\end{align}
where we have highlighted the fact that in $\lambda$, the velocity of the second particle should be evaluated at retarded time.
This Lagrangian shows causal propagation from particle $2$ to particle $1$ and vice-versa. The symmetric term $(1 \leftrightarrow 2)$ can also be rewritten as a dependence on advanced time, as in the Feynman-Wheeler absorber theory \cite{Wheeler:1945aa}.

This Lagrangian has a number of interesting properties. First, it is exactly Poincaré invariant, as we have integrated out gravitons without breaking Poincaré invariance. At the level of the effective Lagrangian, while the invariance under spacetime translations and space rotations is obvious, it is not immediately clear that $L$ is also invariant under Lorentz boosts. But remember that the combination of velocities in $\lambda$ comes from the Lorentz invariant contraction $P_{\mu \nu; \alpha \beta} v_1^\mu v_1^\nu v_2^\alpha v_2^\beta$, while the Liénard-Wiechert dependence on the positions is also Lorentz invariant. Contrast this with the usual post-Newtonian Lagrangian, which is Lorentz invariant only to next order in the PN expansion.

This exact Poincaré invariance implies the conservation of the ten usual quantities : momentum, angular momentum, energy and center-of-mass theorem. We will use these conserved quantities to derive the equations of motion of circular orbits in Section \ref{sec:circular}.

Another particularity of the Lagrangian \eqref{eq:my_eq_retarded} (which in fact is just a consequence of its Poincaré invariance) is that it is conservative, i.e it represents a system which does not dissipate any energy in the form of gravitational waves. Of course, such a system is perfectly unphysical, but conservative Lagrangians are nonetheless useful because there is a clean separation between conservative and dissipative dynamics in most approaches to the two-body problem \cite{Blanchet:2013haa}. Some properties of the full dynamics are due to its conservative part (such as the existence of an innermost circular orbit), and other properties emerge from the dissipative part (such as the adiabatic inspiral of the two point-particles). In this article we will only focus on conservative dynamics.
 
Finally, it is easy to show that by integrating out the einbeins $e_\alpha$ in a PM expansion (i.e, an expansion in powers of $G$), one recovers the 1PM Lagrangian presented in \cite{Foffa:2013gja}.

\section{Post-Newtonian expansion} \label{sec:PN_expansion}

\subsection{Instantaneous dynamics} 

Let us now focus on the instantaneous Lagrangian \eqref{eq:my_Lagrangian} and show how we recover the standard post-Newtonian expansion. To get the effective two-body dynamics from the Lagrangian \eqref{eq:my_Lagrangian}, one should integrate out the two auxiliary parameters. This gives the two equations
\begin{align} \label{eq:system_e1e2}
\begin{split}
e_1^2 &= 1 - v_1^2 - \frac{2 \lambda G m_2}{e_2 r} \; , \\
e_2^2 &= 1 - v_2^2 - \frac{2 \lambda G m_1}{e_1 r} \; ,
\end{split}
\end{align}
which together imply a fifth-order equation for e.g $e_1$,
\begin{equation}
(e_1^2 - 1 + v_1^2)^2 \left( e_1(1-v_2^2) - \frac{2 \lambda G m_1}{r} \right) - \frac{4 \lambda^2 G^2 m_2^2 e_1 }{r^2} = 0  \; ,
\end{equation}
which will be central in what follows. By applying the post-Newtonian scaling $1/r = \mathcal{O}(v^2) \ll 1$ it is easy to perturbatively solve this equation. There are four branches of solutions and we select the one whose PN expansion is consistent. We find
\begin{align}
\begin{split} \label{eq:PN_sol_e1}
e_1 &= 1 - \frac{1}{2} \left(v_1^2 + \frac{2 G m_2}{r} \right) \\
&- \frac{1}{8} \left(v_1^4 +  \frac{G m_2}{r}( 12 v_1^2 + 12 v_2^2 - 32 \mathbf{v}_1 \cdot \mathbf{v}_2 ) + \frac{4 G^2 m_2(m_2 + 2 m_1)}{r^2}  \right) + \mathcal{O}(v^6) \; ,
\end{split}
\end{align}
which gives the two-body Lagrangian up to order $\mathcal{O}(v^6)$,
\begin{align}
\begin{split}
L &= -(m_1+m_2) + \frac{1}{2} \left(m_1 v_1^2 + m_2 v_2^2 + \frac{2 G m_1 m_2}{r} \right) \\
&+ \frac{1}{8} \left( m_1 v_1^4 + m_2 v_2^4 + 4 \frac{G m_1 m_2}{r} \left(3 v_1^2 + 3v_2^2 - 8 \mathbf{v}_1 \cdot \mathbf{v}_2  \right) + \frac{4 G^2 m_1 m_2(m_1 + m_2)}{r^2} \right) + \mathcal{O}(v^6) \; .
\end{split}
\end{align}

In the $\mathcal{O}(v^2)$ term we recognize the usual Newtonian potential. The next order should give the 1PN or Einstein-Infeld-Hoffmann Lagrangian. However, recall that we do not yet consider \textit{bulk} nonlinearities, as well as propagators corrections coming from retarded effects. This means that our expression should only recover diagrams 4b, 4c and 5b of Ref. \cite{goldberger_effective_2006}, and it is indeed the case. It is remarkable that by a single linear diagram one can get directly the diagrams with nonlinear wordline fields insertions such as diagram 5b of Ref. \cite{goldberger_effective_2006}. Once propagator insertions are included, to obtain the full 1PN Lagrangian we will only miss one diagram which is the one with the cubic graviton vertex given by Figure 5b in Ref. \cite{goldberger_effective_2006}.

Going further, one may want to check if this property holds also at the 2PN order. Continuing the procedure outlined before, we find
\begin{align}
\begin{split}\label{eq:L2PN}
L_{2PN} &= \frac{1}{16} \left(m_1 v_1^6 + m_2 v_2^6 + 2 \frac{G m_1 m_2}{r} \left(7 v_1^4 + 7 v_2^4 + 2 v_1^2 v_2^2 - 16 \mathbf{v}_1 \cdot \mathbf{v}_2 (v_1^2 + v_2^2)  + 16 (\mathbf{v}_1 \cdot \mathbf{v}_2)^2 \right)  \right. \\
&+ \left. 4 \frac{G^2 m_1 m_2}{r^2} \left(v_1^2(6m_1+7m_2) + v_2^2(6m_2+7m_1) - 16 \mathbf{v}_1 \cdot \mathbf{v}_2 (m_1+m_2)  \right) + 8 \frac{G^3 m_1 m_2(m_1+m_2)^2}{r^3}  \right) \; .
\end{split}
\end{align}

The 2PN conservative Lagrangian in the NRGR formalism has been derived in Ref. \cite{Gilmore_2008}. However, when comparing our results one should be careful about the fact that Ref. \cite{Gilmore_2008} uses a different parametrization of the metric, namely the Kol-Smolkin variables \cite{Kol:2007rx},
\begin{equation}
g_{\mu \nu} = \begin{pmatrix}
e^{2 \phi / \mpl} & - e^{2 \phi / \mpl} A_j / \mpl \\
- e^{2 \phi / \mpl} A_i / \mpl \vspace{1em} & - e^{-2 \phi / \mpl} \gamma_{ij} + e^{2 \phi / \mpl} A_i A_j / \mpl^2
\end{pmatrix} \; ,
\end{equation}
where the metric excitations are described in terms of one scalar $\phi$, one vector $A_i$ and one tensor $\gamma_{ij}$. Thus, our perturbation $h_{\mu \nu}$ contains in fact an infinite number of powers of $\phi$. This means that our linearized Einstein-Hilbert action \ref{eq:EH_second_order} contains an infinite number of bulk vertices in the Kol-Smolkin variables (but not all of them). This makes a direct comparison with Ref. \cite{Gilmore_2008} difficult at the nonlinear level. 

In the following we will separate the radial and velocity dependence of the different terms at a given PN level. This means that at 2PN order, there are four classes of terms entering eq. \eqref{eq:L2PN} : $\mathcal{O}(v^6)$ (trivial since it just comes from the expansion of $\sqrt{1-v^2}$), $\mathcal{O}(v^4/r)$, $\mathcal{O}(v^2/r^2)$ and $\mathcal{O}(1/r^3)$. The diagrams giving the $\mathcal{O}(v^4/r)$ terms do not involve any bulk vertices, so that we can directly compare our results to those of Ref. \cite{Gilmore_2008} which are contained in diagrams $a$, $d$ and $f$ of this reference. Indeed, the sum of these diagrams exactly gives the first line of eq. \eqref{eq:L2PN}.

\begin{figure}
	\centering
	\subfloat[]{ \label{subfig:2PN_a}
%
%
%
%
\includegraphics[scale=1]{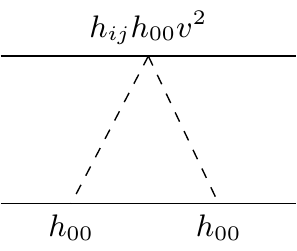}
	} \hspace{1em}
	\subfloat[]{ \label{subfig:2PN_b}
%
%
%
%
\includegraphics[scale=1]{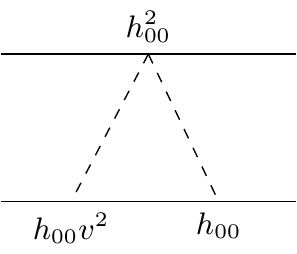}
	} \hspace{1em}
	\subfloat[]{ \label{subfig:2PN_c}
%
%
%
%
\includegraphics[scale=1]{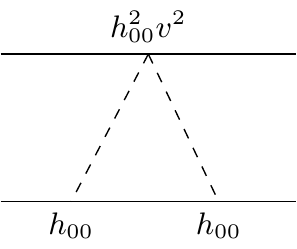}
	} \hspace{1em}
	\subfloat[]{ \label{subfig:2PN_d}
%
%
%
%
\includegraphics[scale=1]{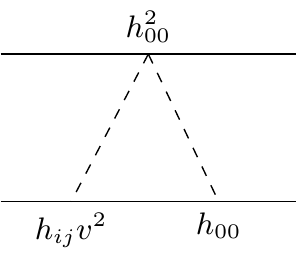}
	} \hspace{1em}
	\subfloat[]{ \label{subfig:2PN_e}
%
%
%
%
\includegraphics[scale=1]{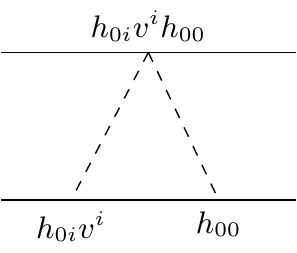}
	} \hspace{1em}
	\subfloat[]{ \label{subfig:2PN_f}
%
%
%
%
\includegraphics[scale=1]{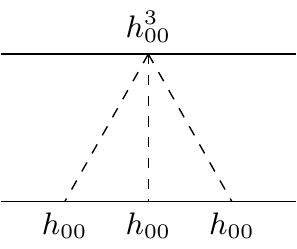}
	} \hspace{1em}
	\subfloat[]{ \label{subfig:2PN_g}
%
%
%
%
\includegraphics[scale=1]{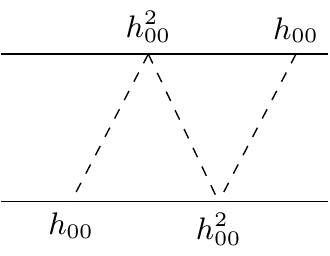}
	} \hspace{1em}

\caption{The seven Feynman diagrams needed at 2PN order. Each vertex is labeled with the corresponding term it refers to in the expansion of the square root in the matter action \ref{eq:pp_action}. Each diagram which is not symmetric under the exchange $1 \leftrightarrow 2$ should be added with its symmetric counterpart.}
\label{fig:Feyn_2PN}
\end{figure}

Concerning the second line of eq. \eqref{eq:L2PN}, we will rather evaluate ourselves the diagrams needed without using the Kol-Smolkin variables for the reason explained above. The diagrams needed to be evaluated are represented in Figure \ref{fig:Feyn_2PN} : since they do not involve bulk nonlinearities, they are rather straightforward to evaluate using the rules of NRGR. We find

\begin{align}
\mathrm{Fig} \; \ref{subfig:2PN_a} &= i \frac{G^2 m_1 m_2^2}{r^2} v_1^2 + (1 \leftrightarrow 2) \; , \\
\mathrm{Fig} \; \ref{subfig:2PN_b} &= i \frac{G^2 m_1 m_2^2}{2 r^2} v_2^2 + (1 \leftrightarrow 2) \; ,  \\
\mathrm{Fig} \; \ref{subfig:2PN_c} &= i \frac{3 G^2 m_1 m_2^2}{4 r^2} v_1^2 + (1 \leftrightarrow 2) \; ,  \\
\mathrm{Fig} \; \ref{subfig:2PN_d} &= i \frac{G^2 m_1 m_2^2}{r^2} v_2^2 + (1 \leftrightarrow 2) \; ,  \\
\mathrm{Fig} \; \ref{subfig:2PN_e} &= - 4 i \frac{G^2 m_1 m_2^2}{r^2} \mathbf{v}_1 \cdot \mathbf{v}_2 + (1 \leftrightarrow 2) \; ,  \\
\mathrm{Fig} \; \ref{subfig:2PN_f} &= i \frac{G^3 m_1 m_2^3}{2 r^3} + (1 \leftrightarrow 2) \; ,   \\
\mathrm{Fig} \; \ref{subfig:2PN_g} &= i \frac{G^3 m_1^2 m_2^2}{r^3} \; . 
\end{align}

Summing all these different contributions, we recover exactly the second line of the 2PN Lagrangian \eqref{eq:L2PN}. However, we also discover that our resummation technique is somewhat not so interesting as one could naively imagine: indeed, all of the diagrams in Figure~\ref{fig:Feyn_2PN}  are quite easy to compute compared to diagrams involving bulk vertices like the one in Figure 5b of Ref. \cite{goldberger_effective_2006}. This means that the true complexity of the GR two-body problem is not contained in the (point-particle) matter sector, but in the gravitational action.

\subsection{Retardation effects}


In this Section we will include the retardation effects in the 1PN and 2PN Lagrangians and show that we also recover known results (in the NRGR formalism, these are taken into account by modifications of the propagator). To do so, it is convenient to rewrite the Feynman propagator as
\begin{equation}
D_F(x_1-x_2) = \frac{- i}{4 \pi} \sum_n \frac{(-1)^n}{(2n)!} \vert \mathbf{x}_1 - \mathbf{x}_2 \vert^{2n-1} \frac{\mathrm{d}^{2n}}{\mathrm{d}t_1^n \mathrm{d}t_2^n} \delta(t_1-t_2) \; ,
\end{equation}
where we have expanded the delta functions in term of $\vert \mathbf{x}_1 - \mathbf{x}_2 \vert$. This shows that the interacting action can be rewritten as
\begin{equation} \label{eq:Sint_powerSeries}
S_\mathrm{int} = \sum_n \frac{(-1)^n}{(2n)!} \int \mathrm{d} t \frac{\mathrm{d}^{2n}}{\mathrm{d}t_1^n \mathrm{d}t_2^n} \left[ \frac{G m_1 m_2 \lambda(t_1, t_2) \vert \mathbf{x}_1(t_1) - \mathbf{x}_2(t_2) \vert^{2n-1} }{e_1(t_1) e_2(t_2)} \right]_{t_1 = t_2 = t} \; .
\end{equation}
Note that it is \textit{not} a total derivative as it may seem at first glance. To this interacting action one should also add the 'kinetic' action in the first line of Eq. \eqref{eq:my_eq_retarded}.

At the 1PN level, one can simply set $e_1 = e_2 = \lambda = 1$ and compute the $n=1$ term in eq. \eqref{eq:Sint_powerSeries} (even if $e_1$ or $e_2$ are modified by retardation effects, this dependence will cancel at this order because the 'kinetic' Lagrangian involves the combination $e_\alpha + 1/e_\alpha$). This yields
\begin{equation}
L_{1PN, \mathrm{retarded}} = \frac{G m_1 m_2}{2 r} \left( \mathbf{v}_1 \cdot \mathbf{v}_2 - (\mathbf{v}_1 \cdot \mathbf{n}) (\mathbf{v}_2 \cdot \mathbf{n}) \right) \; ,
\end{equation}
where $\mathbf{n} = \mathbf{x}_1 - \mathbf{x}_2$. This is exactly diagram 4a of Ref. \cite{goldberger_effective_2006}.

At the 2PN level, modifications to the einbeins will need to be taken into account. To this aim, it is convenient to rewrite the interacting action using integration by parts as
\begin{align}
\begin{split} \label{eq:Sint_powerSeries_IPP}
S_\mathrm{int} &= \sum_n \frac{(-1)^n}{(2n)!} \sum_{j,k=0}^n (-1)^{j+k} {n \choose j} {n \choose k} \int \mathrm{d} t \frac{1}{e_1(t) e_2(t)} \\
& \times \frac{\mathrm{d}^{j+k}}{\mathrm{d} t^{j+k}} \left[ \left. \frac{\mathrm{d}^{2n-j-k}}{\mathrm{d} t_1^{n-j} \mathrm{d} t_2^{n-k} } \left[ G m_1 m_2 \lambda(t_1, t_2) \vert \mathbf{x}_1(t_1) - \mathbf{x}_2(t_2) \vert^{2n-1} \right] \right\vert_{t_1 = t_2 = t} \right] \; ,
\end{split}
\end{align}
which allows for a direct variation of the action with respect to $e_1$ and $e_2$.

There are now three contributions from retardation effects to the effective action at the 2PN order : the $n=2$ term in eq. \eqref{eq:Sint_powerSeries} taking $e_1 = e_2 = \lambda = 1$, the $n=1$ term in the same equation with the 1PN correction to $e_1$, $e_2$ and $\lambda$, and the 2PN term in the kinetic Lagrangian coming from the 1PN correction to $e_1$ and $e_2$. We consider here only corrections to the $\mathcal{O}(v^4/r)$ sector of the Lagrangian to allow for a direct comparison with Ref. \cite{Gilmore_2008}, as explained in the previous Section. Taking them all into account recovers exactly diagrams $b$, $c$ and $e$ of this reference. 

\subsection{Finite-size effects}

A great improvement of our approach compared to standard NRGR is that finite-size effect have a more direct graphical interpretation : they are the only non-minimal couplings to the point-particles worldlines. These effects are known to arise at the 5PN order for nonspinning sources and consequently necessitate the knowledge of the PN dynamics up to this very high order. Following \cite{porto_effective_2016}, we can model finite-size effects at lowest order by adding a term in the action
\begin{equation} \label{eq:CE}
\frac{C_E}{\mpl^2} \int \mathrm{d}t (\partial_i \partial_j h_{00})^2 \; ,
\end{equation}
where the coefficient $C_E$ is related to Love numbers with a scaling $C_E \sim r_s^5/G$ with $r_s$ the size of the source. We choose here to treat these kind of operators perturbatively, while still staying nonperturbative in the lowest-order point-particle coupling $-m \int \mathrm{d} \tau$. Hence a simple Feynman diagram gives the lowest-order contribution of finite-size effects to the effective Lagrangian \eqref{eq:my_eq_retarded} :
\begin{figure}
	\centering
	\subfloat[]{
%
%
%
%
\includegraphics[scale=1]{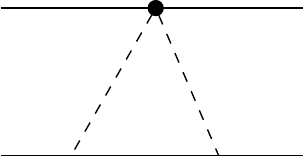}
	} \hspace{1em}

\caption{Leading order contribution to the finite-size effects. The dotted line represents an insertion of the nonminimal coupling operator \ref{eq:CE}. The diagram should be added with its symmetric counterpart.}
\label{fig:Feyn_finite-size}
\end{figure}
\begin{equation}
\mathrm{Figure} \ref{fig:Feyn_finite-size} = 24 i \frac{G^2 C_{E,1} m_2^2}{e_2^2 r^6} + (1 \leftrightarrow 2) \; .
\end{equation}


\section{Static potential} \label{sec:static_pot}

Having confidently established the perturbative validity of our Lagrangian \eqref{eq:my_eq_retarded}, we will now analyze in more details the non-perturbative effects it implies. By non-perturbative we mean that we do not resort to any post-Newtonian expansion beyond the first approximation of ignoring bulk nonlinearities. This approach is similar in spirit to Ref. \cite{Antonelli:2019ytb}, who choose to consider PM hamiltonians as exact in order to gauge the improvement of the PM approximation when comparing to a PN expansion. Of course, there is no physically conceivable situation where the bulk nonlinearities would be subdominant while not being in a post-Newtonian approximation scheme. However, as stated in the Introduction, our results will be in some sense 'more exact' than the post-Newtonian ones. We will see that we will be able to define an 'effective two-body horizon', a quantity whose very existence cannot be recasted in the standard post-Newtonian formalism.

 To get some insight, we will begin by analyzing the static potential between the two point-particles. This means that we will set $v_1 = v_2 = 0$ in all our preceding formulas and imagine that an external operator pinpoints the two masses at their location so that they do not move (of course, such a procedure is perfectly unphysical ; we will deal with measurable quantities in Section \ref{sec:circular}). The potential energy of the system is simply the opposite of the Lagrangian and is given by
\begin{equation}
V(r) =  \frac{m_1}{2} \left( e_1 + \frac{1}{e_1} \right) + \frac{m_2}{2} \left( e_2 + \frac{1}{e_2} \right) - \frac{G m_1 m_2}{e_1 e_2 r} \; ,
\end{equation}
and $e_1, e_2$ obey the quintic equation
\begin{equation} \label{eq:quintic_static}
f_1(e_1, r) \equiv (e_1^2 - 1)^2 \left( e_1 - \frac{2 G m_1}{r} \right) - \frac{4 G^2 m_2^2 e_1 }{r^2} = 0 \; ,
\end{equation}
with labels interchanged for $e_2$. 

\begin{figure}
\center \includegraphics[width=.6\columnwidth]{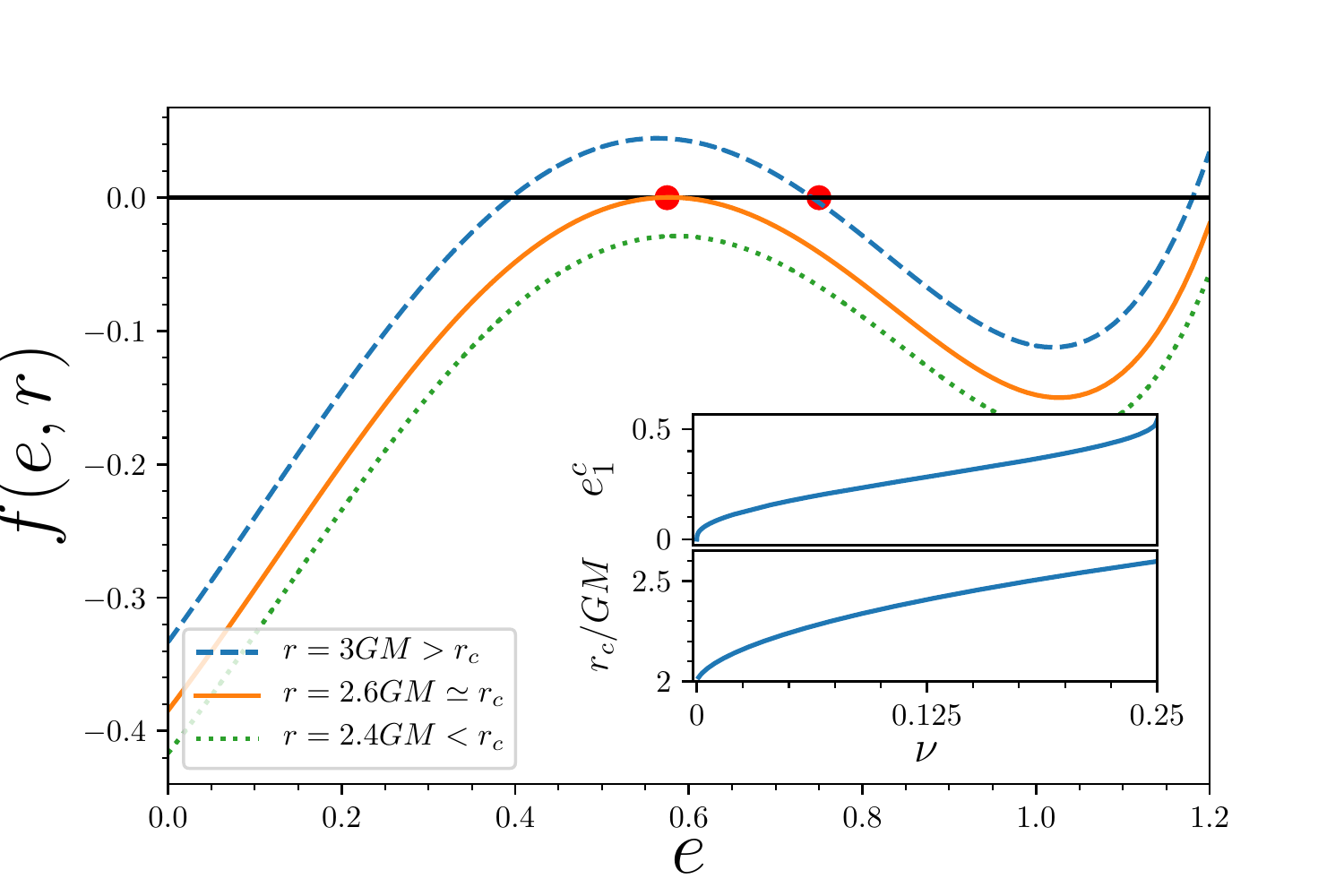}
\caption{\textit{Main plot}: Polynomial equation \eqref{eq:quintic_static} in the equal-mass case $m_1 = m_2 = M/2$, for different values of $r$. The red dot represents the solution with the correct post-Newtonian behavior for $r \rightarrow \infty$. For $r$ smaller than the critical radius $r_c$ this solution becomes complex-valued.
\textit{Subplot}: Critical radius in units of $GM$ and critical parameter $e_1$, as a function of the symmetric mass ratio $\nu = m_1 m_2/M^2$. It is assumed that $m_1 < m_2$, so that in the test-mass limit $m_1 \rightarrow 0$, $r_c = 2G M = 2Gm_2$ and $e_1^c = 0$; in the equal-mass case, $r_c = 3 \sqrt{3} G M /2 \simeq 2.6 G M$ and $e_1^c = 1/\sqrt{3} \simeq 0.58$ }
\label{fig:quintic_static}
\end{figure}

The polynomial equation \eqref{eq:quintic_static} has five roots and only one has the correct post-Newtonian expansion as in \eqref{eq:PN_sol_e1}.
In Figure \ref{fig:quintic_static} we have plotted the quintic equation \eqref{eq:quintic_static} in the equal-mass case as a function of $e_1$ and for different values of $r$. It appears that for a certain critical value of the radius $r_c$, the root $e_1(r)$ with the correct post-Newtonian behavior cease to exist (more precisely, this root becomes complex). This critical radius is defined as the point where
\begin{equation}
f_1(e_1, r) = \frac{\partial f_1}{\partial e_1} = 0 \; ,
\end{equation} 
and is also plotted as a function of the symmetric mass ratio $\nu = m_1 m_2/(m_1+m_2)^2$ in Figure \ref{fig:quintic_static} (an exact expression for $r_c$ does exist but is not so illuminating). Note that due to the symmetry of the equations, the critical radius is the same if one instead considers the polynomial equation on $e_2$.

To get an understanding of the phenomenon at play, let us consider the test-mass ratio $m_1 \rightarrow 0$. In this case the critical radius becomes $r=2Gm_2$. But we also know the exact solution for $e_1$ from the original point-particle action \eqref{eq:pp_action_e} : it is $e_1^2 = -\bar g_{00}$  where $\bar g_{\mu \nu}$ is the \sch metric in harmonic coordinates,
\begin{equation} \label{eq:g00_harmonic}
\bar g_{00} = - \frac{r-Gm_2}{r+Gm_2} \; .
\end{equation}
More precisely, since we are considering a one-graviton exchange in the Lagrangian \eqref{eq:my_eq_retarded} we should rather take the linearized \sch metric so that $e_1^2 \simeq 1 - 2Gm_2/r$.
 It is then clear that in this case the critical radius corresponds to the point-particle becoming lightlike $e_1=0$, i.e the critical radius is reached when the point-particle is on the horizon of the massive particle $m_2$.
In this sense, our equations define an effective horizon for two interacting point-particles : for $r < r_c$, our static assumption is necessarily invalid since there are no solution to the equations, i.e the two particles are forced to move. We will translate this statement in a measurable gauge-invariant quantity in the Section \ref{sec:circular}.

Of course, the true horizon of the \sch metric in harmonic coordinates is situated at $r=Gm_2$ and not at $r=2Gm_2$.
 We expect that taking into account cubic and higher vertices would give a more accurate estimate of the location of the horizon. This is indeed the case when approximating the rational fraction \eqref{eq:g00_harmonic} by polynomials of increasing degree in order to find its zero.

It is interesting to note that the disappearance of a solution of the quintic equation \eqref{eq:quintic_static} is an intrinsically nonperturbative phenomenon. Indeed, if one tried to solve the quintic equation for $e_1$ \eqref{eq:quintic_static} perturbatively as in Section \ref{sec:PN_expansion}, one would get e.g
\begin{equation} \label{eq:PN_static}
e_1 = 1 - \frac{G m_2}{r} - \frac{G^2m_2(m_2+2m_1)}{2r^2} - \frac{G^3 m_2(m_2^2 + 4m_1 m_2 + 3 m_1^2)}{2r^3} + \mathcal{O}\left( \frac{1}{r^4} \right) \; ,
\end{equation}
and this solution exists for all $r$. Likewise, one could solve equation \eqref{eq:quintic_static} perturbatively in the mass ratio $m_1/m_2$ to get
\begin{equation}
e_1 = \sqrt{1 - \frac{2Gm_2}{r}} - \frac{G^2 m_1 m_2}{r(r-2Gm_2)} - \frac{3 G^3 m_1^2 m_2(r-Gm_2)}{2r^{3/2}(r-2Gm_2)^{5/2}} + \mathcal{O}(m_1^3) \; .
\end{equation}
The point is that, as the lowest-order solution presents a singular behavior only at $r=2Gm_2$, the same will be true of any perturbative order beyond the leading one, and so the solution for $e_1$ can never cease to exist for any $r$ greater than $2Gm_2$ contrary to the behavior identified in Figure \ref{fig:quintic_static}.

Another feature of the critical point is the characteristic behavior of $e$ near to this point. One can define a 'critical exponent' $\gamma$ by the scaling $e_1 - e_1^c \propto (r - r_c)^\gamma$ close to the critical point, with $\gamma = 1/2$. This is because the vanishing of the derivative of the polynomial equation $\partial f / \partial e_1 = 0$ at the critical point imposes that for $e_1$ close to $e_1^c$ and $r$ close to $r_c$,
\begin{equation}
0 = f_1(e_1, r) \simeq  (r-r_c) \frac{\partial f_1}{\partial r} + \frac{1}{2} (e_1-e_1^c)^2 \frac{\partial^2 f_1}{\partial e_1^2} \; ,
\end{equation}
such that $e_1 - e_1^c \propto (r - r_c)^{1/2}$.

We now move on to the case of circular orbits.

\section{Circular orbits} \label{sec:circular}

In this Section, we will derive the perturbative and nonperturbative properties of the conservative Lagrangian \eqref{eq:my_eq_retarded} (which we simply denote by $L$) in the case of an exactly circular two-body orbit of frequency $\omega$. In this setting, a physical interpretation of the auxiliary parameters $e_1, e_2$ has been given by Detweiler \cite{Detweiler:2008ft}. Namely, these are the redshift of photons emitted near to the point-particles and detected by an observer situated far from the system on its axis of rotation, and as such are gauge invariant within the physically reasonable class of gauges defined in \cite{Blanchet:2011aha}. In the following, we will mainly be interested by this \textit{redshift observable}, as well as by the energy of circular orbits, considered as functions of the frequency of the orbit.

\subsection{Equations of motion}

In order to derive the equations of motion of e.g the first particle, we choose to rewrite the $(1 \leftrightarrow 2)$ term in \eqref{eq:my_eq_retarded} as a dependence on advanced time for the particle 2, so that the variables concerning the first particle are always evaluated at time $t$:
\begin{align}
\begin{split} \label{eq:Lagrangian_advanced}
L &=  - \frac{m_1}{2} \left(e_1(t) + \frac{1-v_1^2}{e_1(t)} \right) - \frac{m_2}{2} \left(e_2(t) + \frac{1-v_2^2}{e_2(t)} \right) 
 + \frac{G m_1 m_2}{2 e_1(t)} \left[  \frac{\lambda(t, t^R) }{ e_2(t^R) \tilde r^R} 
 + \frac{\lambda(t, t^A) }{ e_2(t^A) \tilde r^A} \right] \; ,
\end{split}
\end{align}
where
\begin{align} \label{eq:def_retarded_advanced}
\begin{split}
t^R &= t - \left \vert \mathbf{x}_1(t) - \mathbf{x}_2(t^R) \right \vert \; , \\
t^A &= t + \left \vert \mathbf{x}_1(t) - \mathbf{x}_2(t^A) \right \vert \; , \\
\tilde r^R &= \left \vert \mathbf{x}_1(t) - \mathbf{x}_2(t^R) \right \vert - \mathbf{v}_2(t^R) \cdot (\mathbf{x}_1(t) - \mathbf{x}_2(t^R)) \; , \\
\tilde r^A &= \left \vert \mathbf{x}_1(t) - \mathbf{x}_2(t^A) \right \vert + \mathbf{v}_2(t^A) \cdot (\mathbf{x}_1(t) - \mathbf{x}_2(t^A)) \; .
\end{split}
\end{align}

As stated below Eq. \eqref{eq:my_eq_retarded}, exact Poincaré invariance of the Lagrangian implies the conservation of the ten usual quantities. In particular, conservation of angular momentum restricts the motion to a two-dimensional plane.
We thus parameterize the circular trajectories according to
\begin{equation} \label{eq:trajectory_x1x2}
\mathbf{x}_1(t) = R_1 (\cos \omega t, \sin \omega t)^T \; , \quad \mathbf{x}_2(t) = - R_2 (\cos \omega t, \sin \omega t)^T \; ,
\end{equation}
where $\omega$ is the frequency of the circular orbit. This ansatz solves both the equations of motion and the center-of-mass theorem (see below) ; indeed, because of the time-symmetric (non-dissipative) character of the equations, the momentum of each particle will be aligned with the common axis of the velocities.

By defining the two (positive) variables
\begin{equation}
u^R = t - t^R \; , \quad u^A = t^A - t \; ,
\end{equation}
and plugging the trajectory \eqref{eq:trajectory_x1x2} into the definitions of the retarded and advanced times \eqref{eq:def_retarded_advanced}, we find that $u^R= u^A \equiv u$ satisfy the same equation,
\begin{equation} \label{eq:retarded}
u = \sqrt{R_1^2 + R_2^2 + 2R_1 R_2 \cos \omega u} \; .
\end{equation}

In principle there could be multiple solutions to this equation, but we can focus on the one continuously related to the nonrelativistic solution at large distances ($\omega \rightarrow 0$) where $u = R_1 + R_2$.


Then, the equality $u^R= u^A $ implies that $\tilde r^R$ and $\tilde r^A$ take the common value
\begin{equation}
\tilde r^R = \tilde r^A \equiv \tilde r = u + \omega R_1 R_2 \sin \omega u \; ,
\end{equation}
and similarly for $\lambda$,
\begin{equation}
\lambda(t, t^R) = \lambda(t, t^A) \equiv \tilde \lambda = 1 + \omega^2(R_1^2 + R_2^2 + 4 R_1 R_2 \cos \omega u) + R_1^2 R_2^2 \omega^4 \cos 2 \omega u \; .
\end{equation}

We are now in position to compute the polynomial equations on $e_i$. To this aim, we will rather use the Lagrangian expanded in a power series \eqref{eq:Sint_powerSeries_IPP}. Minimization with respect to $e_1$ gives
\begin{equation}
e_1^2 = 1 - v_1^2 - \frac{2Gm_2}{e_2} \left( \sum_n \frac{(-1)^n}{(2n)!} \sum_{j,k=0}^n (-1)^{j+k} {n \choose j} {n \choose k}  \frac{\mathrm{d}^{j+k}}{\mathrm{d} t^{j+k}} \left[ \left. \frac{\mathrm{d}^{2n-j-k}}{\mathrm{d} t_1^{n-j} \mathrm{d} t_2^{n-k} } \left[ \lambda(t_1, t_2) \vert \mathbf{x}_1(t_1) - \mathbf{x}_2(t_2) \vert^{2n-1} \right] \right\vert_{t_1 = t_2 = t} \right] \right) \; .
\end{equation}
However, since the $(t_1,t_2)$ dependence in $\lambda(t_1, t_2)$ and $\vert \mathbf{x}_1(t_1) - \mathbf{x}_2(t_2) \vert$ is only contained in terms proportional to $\cos \omega(t_1-t_2)$, this equation simplifies drastically. Indeed, by setting $t_1=t_2=t$ the expressions become constant in time, so the derivation with respect to $t$ selects the term $j=k=0$ only. The equation simplifies to
\begin{equation}
e_1^2 = 1 - v_1^2 - \frac{2Gm_2}{e_2} \left( \sum_n \frac{(-1)^n}{(2n)!} \frac{\mathrm{d}^{2n}}{\mathrm{d} t_1^{n} \mathrm{d} t_2^{n} } \left[ \lambda(t_1, t_2) \vert \mathbf{x}_1(t_1) - \mathbf{x}_2(t_2) \vert^{2n-1} \right] \right) \; .
\end{equation}
One easily recognizes the usual expansion in term of retarded and advanced times (trading the derivatives on $t_1$ to derivatives on $t_2$ by using the time dependence $\propto \cos \omega(t_1-t_2)$) , and so the equation is
\begin{equation}
e_1^2 = 1 - v_1^2 - \frac{Gm_2}{e_2} \left( \frac{\lambda(t,t^R)}{\tilde r^R} + \frac{\lambda(t,t^A)}{\tilde r^A} \right) \; .
\end{equation}
With our previous notations, this becomes
\begin{equation}
e_1^2 = 1 - R_1^2 \omega^2 - \frac{2 \tilde \lambda G m_2}{\tilde r e_2}  \; .
\end{equation}
The equation on $e_2$ can simply be found by interchanging $1 \leftrightarrow 2$. The point of this derivation was to show that one can indeed take $e_1$ and $e_2$ to be a constant (independent of time) in all the equations of motion.


Finally, the last equation that we are after is the equation of motion for one of the point-particles (say $m_1$), which will give the generalization of Kepler's law:
\begin{equation} \label{eq:EOM_base}
\frac{d \mathbf{p}_1}{dt} = \frac{\partial L}{\partial \mathbf{x}_1} \; ,
\end{equation}
where 
\begin{equation}
\mathbf{p}_1 = \frac{\partial L}{\partial \mathbf{v}_1} = \frac{m_1}{e_1} \mathbf{v}_1 + \frac{G m_1 m_2}{2 e_1 e_2 \tilde r} \left( \frac{\partial \lambda(t,t^R)}{\partial \mathbf{v}_1} + \frac{\partial \lambda(t,t^A)}{\partial \mathbf{v}_1} \right) \; .
\end{equation}
which upon using the ansatz \eqref{eq:trajectory_x1x2} on $\mathbf{x}_1, \mathbf{x}_2$ gives
\begin{equation}
\mathbf{p}_1 = \bigg[ \frac{m_1}{e_1} R_1 \omega + \frac{G m_1 m_2}{e_1 e_2 \tilde r} \left( 2(1-R_2^2 \omega^2) R_1 \omega + 4 \cos \omega u (1+R_1 R_2 \omega^2 \cos \omega u) R_2 \omega \right) \bigg] (-\sin \omega t, \cos \omega t)^T  \; .
\end{equation}
Thus, as advertised before, the momentum is aligned with the common direction of the velocities.

 To compute $\partial L / \partial \mathbf{x}_1$, one should be careful to the fact that $t^R$ and $t^A$ depend on $\mathbf{x}_1$, so that 
\begin{equation}
\frac{\partial L}{\partial \mathbf{x}_1} = \frac{G m_1 m_2}{2 e_1 e_2 \tilde r} \left( \frac{\partial \lambda^R}{\partial t^R} \frac{\partial t^R}{\partial \mathbf{x}_1} - \frac{\tilde \lambda}{\tilde r} \frac{\partial \tilde r^R}{\partial \mathbf{x}_1} + (R \leftrightarrow A) \right) \; ,
\end{equation}
where $\lambda^R = \lambda(t, t^R)$. From the definitions of the advanced and retarded times \eqref{eq:def_retarded_advanced} one gets
\begin{align}
\frac{\partial t^R}{\partial \mathbf{x}_1} &= - \frac{\mathbf{r}^R}{\tilde r} \; , \quad  \frac{\partial t^A}{\partial \mathbf{x}_1} =  \frac{\mathbf{r}^A}{\tilde r} \; , \\
\frac{\partial \tilde r^R}{\partial \mathbf{x}_1} &= - \mathbf{v}_2^R + \frac{1 - v_2^2 + \mathbf{a}_2^R \cdot \mathbf{r}^R}{\tilde r} \mathbf{r}^R \; , \quad \frac{\partial \tilde r^A}{\partial \mathbf{x}_1} =  \mathbf{v}_2^A + \frac{1 - v_2^2 + \mathbf{a}_2^A \cdot \mathbf{r}^A}{\tilde r} \mathbf{r}^A \; ,
\end{align}
where $\mathbf{r}^R = \mathbf{x}_1(t) - \mathbf{x}_2(t^R)$, and generically a superscript $R$ denotes evaluation at retarded time (the same being true for $A$). With our parameterization, one gets
\begin{align}
\frac{\partial \lambda^R}{\partial t^R} \frac{\partial t^R}{\partial \mathbf{x}_1} + (R \leftrightarrow A) &= -8 R_1 R_2 \omega^3 \sin \omega u (1 + R_1 R_2 \omega^2 \cos \omega u) \frac{R_1 + R_2 \cos \omega u}{\tilde r} (\cos \omega t, \sin \omega t)^T \\
\frac{\partial \tilde r^R}{\partial \mathbf{x}_1} + (R \leftrightarrow A) &= 2 \left[ R_2 \omega \sin \omega u + (1 + R_1 R_2 \omega^2 \cos \omega u) \frac{R_1 + R_2 \cos \omega u}{\tilde r}  \right] (\cos \omega t, \sin \omega t)^T \; .
\end{align}
Finally, the projection of the equation of motion \eqref{eq:EOM_base} gives the generalized Kepler law,
\begin{align}
\begin{split} \label{eq:EOM_Kepler}
R_1 \omega^2 &= \frac{G m_2}{e_2 \tilde r^2} \left\lbrace \tilde \lambda \left[ R_2 \omega \sin \omega u + (1 + R_1 R_2 \omega^2 \cos \omega u) \frac{R_1 + R_2 \cos \omega u}{\tilde r}  \right] \right. \; , \\
&+ \left. \vphantom{\frac{R_1}{\tilde r}} 4 R_2 \omega^2 (1 + R_1 R_2 \omega^2 \cos \omega u) (R_1^2 \omega \sin \omega u - u \cos \omega u) + 2 (R_2^2 \omega^2 - 1) R_1 \tilde r \omega^2 \right \rbrace \; .
\end{split}
\end{align}

It is easily checked that in the nonrelativistic case $e_2=1, \omega \rightarrow 0, R_1 = m_2/(m_1+m_2)r$ (where $r = R_1 + R_2$), one recovers the usual Kepler law, $\omega^2 r^3 = G (m_1+m_2)$. We now have all the equations needed to solve for the two-body motion, namely : the equation one the retarded time $u$, the definitions of $\tilde r$ and $\tilde \lambda$, the equations of motion and the coupled equations on $e_1,e_2$, which we all rewrite here for convenience:
\begin{align}
\begin{split} \label{eq:system_circular}
u &= \sqrt{R_1^2 + R_2^2 + 2R_1 R_2 \cos \omega u} \; , \\
\tilde r &= u + \omega R_1 R_2 \sin \omega u \; , \\
\tilde \lambda &= 1 + \omega^2(R_1^2 + R_2^2 + 4 R_1 R_2 \cos \omega u) + R_1^2 R_2^2 \omega^4 \cos 2 \omega u \; , \\
R_1 \omega^2 &= \frac{G m_2}{e_2 \tilde r^2} \left\lbrace \tilde \lambda \left[ R_2 \omega \sin \omega u + (1 + R_1 R_2 \omega^2 \cos \omega u) \frac{R_1 + R_2 \cos \omega u}{\tilde r}  \right] \right. \\
&+ \left. \vphantom{\frac{R_1}{\tilde r}} 4 R_2 \omega^2 (1 + R_1 R_2 \omega^2 \cos \omega u) (R_1^2 \omega \sin \omega u - u \cos \omega u) + 2 (R_2^2 \omega^2 - 1) R_1 \tilde r \omega^2 \right \rbrace \; ,
 \\
e_1^2 &= 1 - R_1^2 \omega^2 - \frac{2 \tilde \lambda G m_2}{\tilde r e_2}  \; ,
\end{split}
\end{align}
where we did not write the two other equations on the second point-particle arising from a $1 \leftrightarrow 2$ permutation of the last two equations.

\subsection{Conserved energy}

Once these equations are solved, we can easily obtain the conserved energy as the Hamiltonian of the system. A naive guess for $H$ would be
\begin{equation} \label{eq:guess_H}
H = \mathbf{p}_1 \cdot \mathbf{v}_1 + \mathbf{p}_2 \cdot \mathbf{v}_2 - L \; .
\end{equation}
However, because of the presence of retarded and advanced times, the application of Noether's theorem to time translations is not so straightforward, and in fact we will see that Eq. \eqref{eq:guess_H} is actually incomplete. We will now look in more details at the boundary term in the variation of the action needed for Noether's theorem. We rewrite the action in its particle-symmetric form as
\begin{align}
\begin{split} \label{eq:action_symmetric_noether}
S &= \int_{t_-}^{t_+} \mathrm{d} t \left[ - \frac{m_1}{2} \left(e_1 + \frac{1-v_1^2}{e_1} \right) - \frac{m_2}{2} \left(e_2 + \frac{1-v_2^2}{e_2} \right) +  L_2^R + L_1^R \right] \; ,
\end{split}
\end{align}
where we have introduced boundaries for the integration on the time variable, and the retarded Lagrangians are defined as
\begin{align}
\begin{split}
L_2^R &= \frac{\lambda(t, t_2^R) G m_1 m_2 }{2 e_1 e_2(t_2^R) \left[ \left \vert \mathbf{x}_1 - \mathbf{x}_2(t_2^R) \right \vert - \mathbf{v}_2(t_2^R) \cdot (\mathbf{x}_1 - \mathbf{x}_2(t_2^R))\right]} \; , \\
L_1^R &=  \frac{\lambda(t_1^R, t) G m_1 m_2 }{2 e_1(t_1^R) e_2 \left[ \left \vert \mathbf{x}_1(t_1^R) - \mathbf{x}_2 \right \vert + \mathbf{v}_1(t_1^R) \cdot (\mathbf{x}_1(t_1^R) - \mathbf{x}_2)\right]}  \; , \\
t_1^R &= t - \left \vert \mathbf{x}_1(t_1^R) - \mathbf{x}_2 \right \vert \; , \\
t_2^R &= t - \left \vert \mathbf{x}_1 - \mathbf{x}_2(t_2^R) \right \vert \; . \\
\end{split}
\end{align}

To avoid cluttering notation, in this equation and from now on it is implicit that each variable is evaluated at $t$ when we do not write its argument. Let us consider the variation of this action under an arbitrary transformation $\mathbf{x}_\alpha \rightarrow \mathbf{x}_\alpha + \delta \mathbf{x}_\alpha$. To this aim one should rewrite the action so that it contains only the appropriate variables evaluated at their present time. For the first particle, this can be achieved by the change of variable $t' = t_1^R$ whose Jacobian is
\begin{equation} \label{eq:jacobian}
\frac{\mathrm{d}t}{\left \vert \mathbf{x}_1(t_1^R) - \mathbf{x}_2 \right \vert + \mathbf{v}_1(t_1^R) \cdot (\mathbf{x}_1(t_1^R) - \mathbf{x}_2)} = \frac{\mathrm{d}t'}{\left \vert \mathbf{x}_1(t') - \mathbf{x}_2(t_2^A) \right \vert + \mathbf{v}_2(t_2^A) \cdot (\mathbf{x}_1(t') - \mathbf{x}_2(t_2^A))} \; ,
\end{equation}
with obvious notations for the advanced time of the second particle. This has the effect of transforming $L_1^R$ in $L_2^A$ where the positions of the first particle is evaluated at present time while the second one at advanced time.

 Due to this change of variable, the boundaries change from $t_- \rightarrow t_+$ to $t_{-,1}^R \rightarrow t_{+,1}^R$. Since the boundaries now contain the positions of the particles, we should take them into account when varying the action. The variation of $t_{+,1}^R$ with respect to $\mathbf{x}_1$ is
\begin{equation}
\delta t_{+,1}^R = - \frac{\delta \mathbf{x}_1(t_{+,1}^R) \cdot \big( \mathbf{x}_1(t_{+,1}^R) - \mathbf{x}_2(t_+) \big)}{\left \vert \mathbf{x}_1(t_{+,1}^R) - \mathbf{x}_2(t_+) \right \vert + \mathbf{v}_1(t_{+,1}^R) \cdot (\mathbf{x}_1(t_{+,1}^R) - \mathbf{x}_2(t_+))} \; ,
\end{equation}
and similarly for $t_-$. We can then write the total variation of the action as
\begin{align}
\begin{split}
\delta S &= \int_{t_-}^{t_+} \mathrm{d}t \left\lbrace \frac{\partial L_2^R}{\partial \mathbf{x}_1} \delta \mathbf{x}_1 + \left( \frac{m_1}{e_1} \mathbf{v}_1 + \frac{\partial L_2^R}{\partial \mathbf{v}_1}  \right) \delta \mathbf{v}_1  \right\rbrace \\
&+ \int_{t_{-,1}^R}^{t_{+,1}^R} \mathrm{d}t \left\lbrace \frac{\partial L_2^A}{\partial \mathbf{x}_1} \delta \mathbf{x}_1 + \frac{\partial L_2^A}{\partial \mathbf{v}_1} \delta \mathbf{v}_1   - \frac{\mathrm{d}}{\mathrm{d}t} \left[ L_2^A \frac{\delta \mathbf{x}_1 \cdot \big( \mathbf{x}_1 - \mathbf{x}_2(t_2^A) \big)}{\left \vert \mathbf{x}_1 - \mathbf{x}_2(t_2^A) \right \vert + \mathbf{v}_1 \cdot (\mathbf{x}_1 - \mathbf{x}_2(t_2^A))} \right]  \right\rbrace + (1 \leftrightarrow 2) \; .
\end{split}
\end{align}
Upon integration by parts, and using the equations of motion, $\delta S$ reduces to a boundary term
\begin{align}
\begin{split} \label{eq:deltaS}
\delta S = \int_{t_-}^{t_+} \mathrm{d}t \frac{\mathrm{d}}{\mathrm{d}t} \left[ \mathbf{p}_1 \cdot \delta \mathbf{x}_1 - L_2^A \frac{\delta \mathbf{x}_1(t_1^R) \cdot \big( \mathbf{x}_1(t_1^R) - \mathbf{x}_2 \big)}{\left \vert \mathbf{x}_1(t_1^R) - \mathbf{x}_2 \right \vert + \mathbf{v}_1(t_1^R) \cdot (\mathbf{x}_1(t_1^R) - \mathbf{x}_2)} - F_1  \right] + (1 \leftrightarrow 2) \; ,
\end{split}
\end{align}
where we have used that, when the equations of motion are satisfied, the Jacobian of the change of variable in Eq \eqref{eq:jacobian} is trivial, $\mathrm{d}t = \mathrm{d}t'$. In this equation, the function $F_1$ is defined by
\begin{equation}
F_1(t) = \int_{t_1^R}^t \mathrm{d}t' \left[ \frac{\partial L_2^A}{\partial \mathbf{x}_1} \delta \mathbf{x}_1 + \frac{\partial L_2^A}{\partial \mathbf{v}_1} \delta \mathbf{v}_1 \right] \; .
\end{equation}

We are now ready to derive Nother's theorem. For $\delta \mathbf{x}_1 = \delta \mathbf{x}_2 = \epsilon \mathbf{n}$ with $\epsilon$ a small parameter and $\mathbf{n}$ a constant direction (expressing the invariance of the action under space translations), by setting $\delta S = 0$ we obtain an equation expressing the conservation of total momentum. It can easily be checked to be redundant with the equations of motion \eqref{eq:EOM_Kepler}, and the same is true for the center-of-mass theorem originating from the invariance of the action under boosts.

On the other hand, for $\delta \mathbf{x}_\alpha = \epsilon \mathbf{v}_\alpha$ ($\alpha=1,2$), the variation of the action is a total derivative, $\delta S = \int_{t_-}^{t_+} \mathrm{d}t \; \mathrm{d}L/\mathrm{d}t$ where $L$ is the total Lagrangian. Setting this quantity equal to the one we just computed \eqref{eq:deltaS}, we can express the conservation of the total energy of the system which reads
\begin{align}
\begin{split} \label{eq:energy_circular}
E &= \frac{m_1}{2} \left(e_1+\frac{1+R_1^2 \omega^2}{e_1} \right) + \frac{m_2}{2} \left(e_2+\frac{1+R_2^2 \omega^2}{e_2} \right) \\
& - \frac{Gm_1m_2}{e_1e_2 \tilde r} \bigg[ 1 - \omega^2\left(R_1^2 + R_2^2 + 4 R_1 R_2 \cos \omega u \right)  
  - 3 R_1^2 R_2^2 \omega^4 \cos 2 \omega u \\
  &\quad + 4 \frac{u^2}{\tilde r} R_1 R_2 \omega^3 \sin \omega u \left(1+R_1 R_2 \omega^2 \cos \omega u\right)
+ \left. \frac{\tilde \lambda}{\tilde r} \left( R_1 R_2 \omega \sin \omega u + \frac{u}{\tilde r} R_1 R_2 \omega (\omega u \cos \omega u - \sin \omega u) \right)   \right] \; .
\end{split}
\end{align}

A useful check on the validity of our computation consists in the expansion of the energy as a function of $\omega$ in the post-Newtonian regime, using the scaling $\omega R_\alpha \sim v$, $Gm/R_\alpha \sim v^2$ where $\alpha=1,2$. Finding a perturbative solution of the system of equations \eqref{eq:system_circular} is straightforward. We introduce the standard post-Newtonian parameter
\begin{equation}
x = \left( G M \omega \right)^{2/3} \; ,
\end{equation}
where $M = m_1 + m_2$ is the total mass, in terms of which the total energy $\mathcal{E} = E - m_1 - m_2$  is
\begin{align}
\begin{split}
\mathcal{E} = - \frac{\mu x}{2} \left( 1-\frac{x}{12}  (\nu +17)-\frac{x^2}{24}  \left(\nu ^2-209
   \nu +145\right) -\frac{5 x^3}{5184}  \left(7 \nu ^3+6810 \nu ^2-22593
   \nu +23591\right) + \mathcal{O}(x^4) \right) \; ,
\end{split}
\end{align}
where as usual $\mu = m_1 m_2/(m_1+m_2)$ and $\nu = \mu / (m_1+m_2)$. Although each term in this expansion would need to be corrected to get the correct PN result (our computations are technically of 0PN order), we can observe that the $\nu^n x^n$ coefficient is correct at each order. This was already noticed in Refs. \cite{Kalin:2019rwq,Foffa:2013gja} : the 1PM energy correctly captures the $\nu^n x^n$ coefficient, and since our computation generalizes the 1PM results we happily recover this fact.

\subsection{Innermost circular orbit}

Let us start this Section by considering a point-particle of mass $\mu$ in a circular orbit around a Schwarzschild black hole of mass $M$. It is well-known that the Schwarzschild solution possesses an Innermost Stable Circular Orbit (ISCO) situated at $R = 6 GM$, where $R$ is a gauge-invariant distance defined by
\begin{equation} \label{eq:GI_distance}
R = \left(\frac{GM}{\omega^2} \right)^{1/3} \; .
\end{equation}
This ISCO is situated at the minimum of the energy of the point-particle, which is
\begin{equation}
E_\mathrm{pp} = \mu \left[ \frac{1-2x}{\sqrt{1-3x}} - 1 \right] \; .
\end{equation}
Another feature of Schwarzschild geometry is the existence of a last circular orbit at $R = 3GM$, under which no circular orbits (even unstable) can exist at all. This locus corresponds to the four-velocity $v^\mu$ of the point-particle becoming lightlike thus justifying its name of 'light-ring'.

In the two-body case, various definitions exist for the Innermost Circular Orbit (ICO). One can stick to the minimum of the energy as given by equation \eqref{eq:energy_circular}, however there is no notion of stability in this definition. An analysis of the stability of circular orbits in the post-Newtonian framework can be found in Ref. \cite{Blanchet:2013haa}. In particular, this work suggests that all circular orbits may be stable in the equal-mass case (we recall here that we concentrate on the conservative part of the dynamics, neglecting dissipation which would make the two point-particles fall into each other in a relatively short time).

\begin{figure}
\center \includegraphics[width=.6\columnwidth]{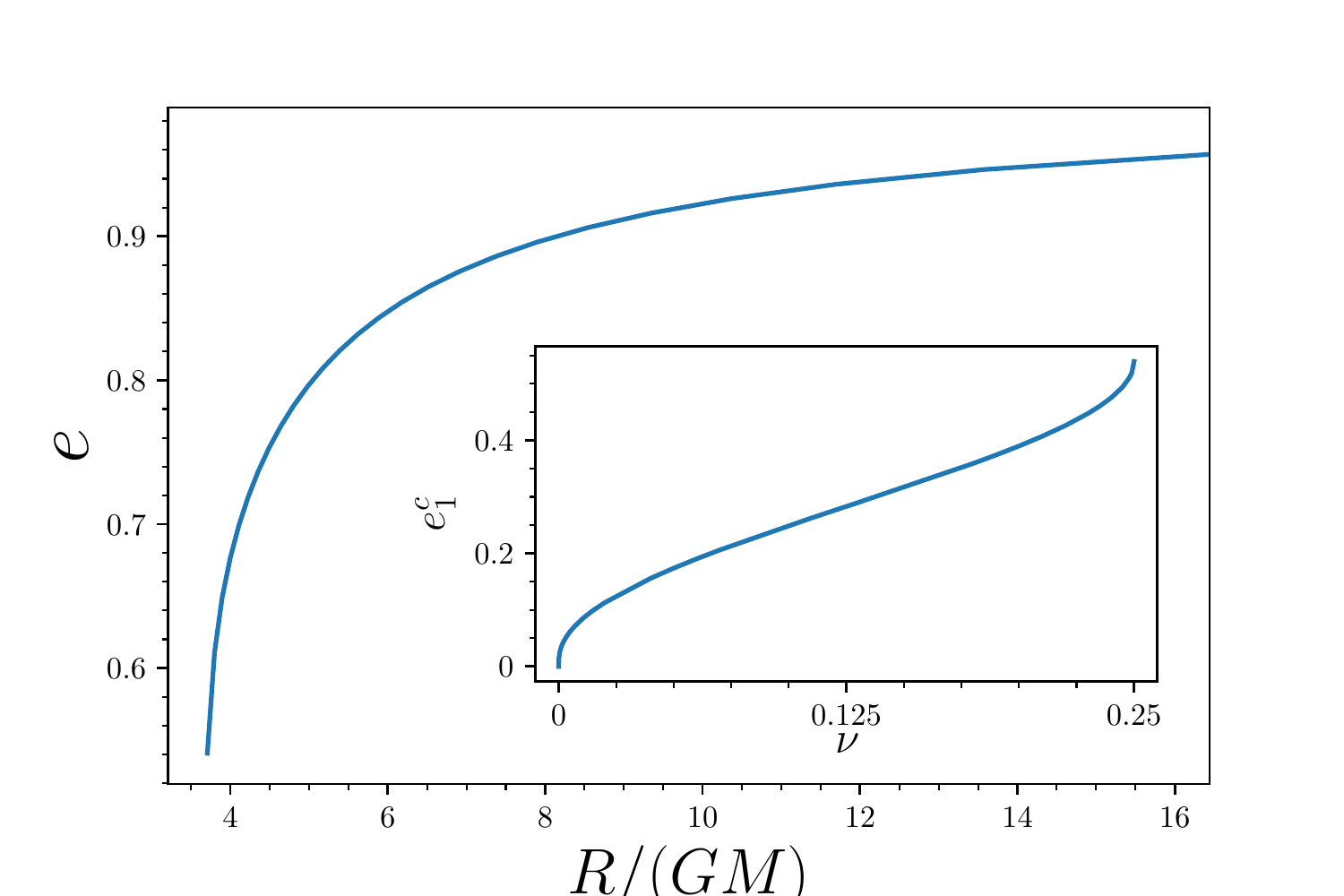}
\caption{\textit{Main plot}: Redshift function $e_1=e_2=e$ in the equal-mass case $m_1=m_2$, plotted as a function of the gauge-invariant distance \eqref{eq:GI_distance}. The derivative of $e$ at the critical radius is infinite.
\textit{Subplot}: Critical redshift $e_1^c$ for different symmetric mass ratios $\nu$. In the test-mass limit, $e_1^c = 0$ ;  in the equal-mass case, $e_1^c \simeq 0.54$.}
\label{fig:ecrit_circular}
\end{figure}

We now add a third (and, we believe, more suited to the name) definition of the innermost circular orbit to the two mentioned above, which we name Critical Innermost Circular Orbit (CICO). It is defined by the point where the redshift functions $e_1, e_2$ become complex-valued \footnote{due to the symmetry of our equations, $e_1$ and $e_2$ become complex-valued at the same value of the frequency $\omega$} : no circular orbit can exist at all beyond the CICO. Moreover, for two particles approaching the CICO an observer situated on the axis of rotation of the binary system would see an abrupt change in the redshift of photons emitted near to the point-particles (i.e., an abrupt change in the functions $e_i$). This is due to the critical behavior observed in Section \ref{sec:static_pot} : at the critical point $(e_c, \omega_c)$, both the polynomial equation on $e_i$ and its derivative vanishes, so that one has the scaling $e - e_c \propto (\omega - \omega_c)^{1/2}$ and the derivative of $e_i$ at the critical point is infinite. This is illustrated in Figure \ref{fig:ecrit_circular}.

\begin{figure}
\center \includegraphics[width=.6\columnwidth]{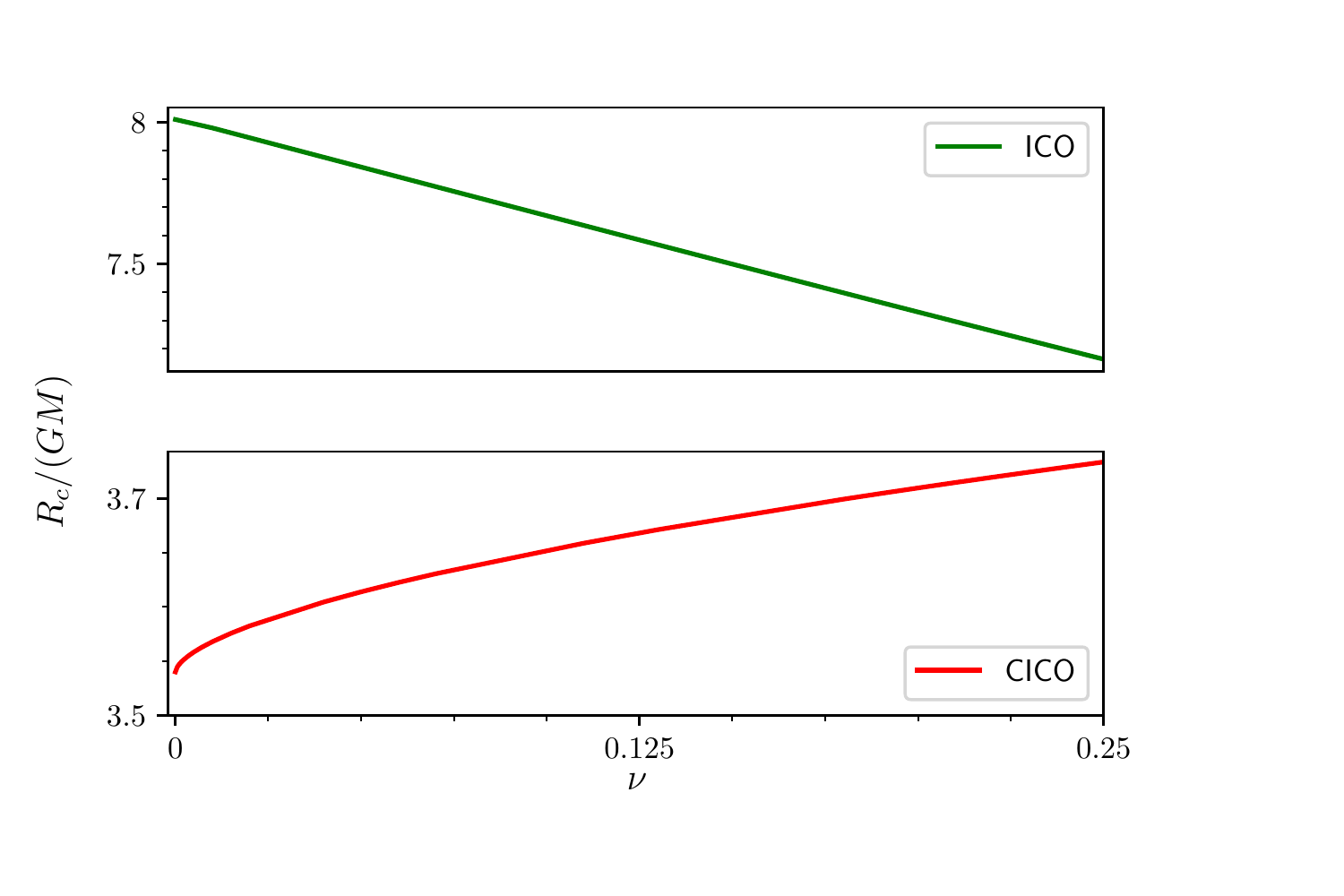}
\caption{Location of the ICO and CICO (translated in the gauge-invariant distance \eqref{eq:GI_distance}) for different symmetric mass ratios $\nu$. The ICO is almost linear in $\nu$. In the test-mass limit, $R^\mathrm{ICO} \simeq 8.01 GM$ and $R^\mathrm{CICO} \simeq 3.54 GM$ ; in the equal-mass case, $R^\mathrm{ICO} \simeq 7.16 GM$ and $R^\mathrm{CICO} \simeq 3.73 GM$. } 
\label{fig:ico_cico}
\end{figure}

Obtaining the value of the ICO by numerically solving the system of equations \eqref{eq:system_circular} is straightforward. In Figure \ref{fig:ico_cico} we have plotted the location of the ICO (given by its 'standard' definition, i.e the minimum of the energy \eqref{eq:energy_circular}) and the CICO as a function of the symmetric mass ratio $\nu$. Note that their value in the test-mass limit $\nu \rightarrow 0$ is not the correct Schwarzschild result since our Lagrangian corresponds to a single graviton exchange. Indeed, our results in the test-mass limit can be recovered by considering the linearized Schwarzschild metric in harmonic coordinates,
\begin{equation}
\mathrm{d}s^2 = - \left(1 - \frac{2GM}{r} \right) \mathrm{d} t^2 + \left(1+\frac{2GM}{r} \right) \left( \mathrm{d} r^2 + r^2 \mathrm{d}\Omega^2 \right) \; .
\end{equation}
In this test-mass limit, the CICO is determined by the equation $g_{\mu \nu} v^\mu v^\nu = 0$, while the ICO is the minimum of the test-mass energy. It is straightforward to derive Kepler's law and the point-particle energy in these coordinates,
\begin{equation}
\omega^2 = \frac{GM}{r^2(r+GM)} \; , \quad E = (r-2GM) \sqrt{\frac{r+GM}{r(r^2-2GMr-4G^2M^2)}} \; .
\end{equation}
Then one readily finds that in the test-mass limit of our linearized approximation the ICO and CICO are situated at
\begin{equation}
R^\mathrm{ICO} \simeq 8.01 GM \; , \quad R^\mathrm{CICO} \simeq 3.54 GM \; .
\end{equation}
Note that the 1PN and 2PN results in the test-mass limit predict respectively $R^\mathrm{ICO, 1PN} \simeq 1.5 GM$ and $R^\mathrm{ICO, 2PN} \simeq 4.02 GM$.

\section{Conclusion}

In this article, we have shown that the introduction of two einbeins allows for a simplification of the Feynman rules of NRGR ; diagrams of increasing complexity are simply recovered from the expansion of a polynomial equation. We thus expect our result to be particularly relevant for the computation of higher order PN dynamics.

Furthermore, we have shown that the polynomial equations obeyed by the redshift variables allows one to define an horizon for two interacting point-particles in GR. For circular orbits, the standard PN solution for the worldline parameters becomes complex-valued for small enough separations so that no circular orbit can exist at all beyond this critical distance. More generally, the disappearance of PN solutions for close enough binaries points towards an inadequacy of the PN parameterization in this strong-field regime.

There are multiple avenues for extending and improving our results. Apart from the inclusion of higher PM orders, it would also be interesting to explore the synergies of our resummation with the Effective One-Body (EOB) \cite{Buonanno:1998gg} formalism. Indeed, while the EOB philosophy is to recast the two-body motion as the one of a point-particle in an effective metric, our resummation is somewhat two-body in essence : worldline nonlinearities do not contribute to the field of an isolated object, so that the Feynman diagrams included in our resummation contain only genuine two-body effects. On the other hand, our treatment misses the one-body dynamics (which is fully contained in the bulk nonlinearities), so that an EOB approach would be complementary.

\begin{acknowledgments}
I would like to thank Filippo Vernizzi, Vitor Cardoso, Scott Melville, Massimiliano Maria Riva, Ira Rothstein and especially Federico Piazza for discussions and comments. I am also grateful to Michele Levi and Rafael Porto for comments on the first version of this paper, and to two referees for their critical reading of the manuscript. This article is partly based upon work from COST Action
GWVerse CA16104, supported by COST
(European Cooperation in Science and Technology).
\end{acknowledgments}

\bibliography{2bodyGR.bib}

\begin{thebibliography}{47}%
\makeatletter
\providecommand \@ifxundefined [1]{%
 \@ifx{#1\undefined}
}%
\providecommand \@ifnum [1]{%
 \ifnum #1\expandafter \@firstoftwo
 \else \expandafter \@secondoftwo
 \fi
}%
\providecommand \@ifx [1]{%
 \ifx #1\expandafter \@firstoftwo
 \else \expandafter \@secondoftwo
 \fi
}%
\providecommand \natexlab [1]{#1}%
\providecommand \enquote  [1]{``#1''}%
\providecommand \bibnamefont  [1]{#1}%
\providecommand \bibfnamefont [1]{#1}%
\providecommand \citenamefont [1]{#1}%
\providecommand \href@noop [0]{\@secondoftwo}%
\providecommand \href [0]{\begingroup \@sanitize@url \@href}%
\providecommand \@href[1]{\@@startlink{#1}\@@href}%
\providecommand \@@href[1]{\endgroup#1\@@endlink}%
\providecommand \@sanitize@url [0]{\catcode `\\12\catcode `\$12\catcode
  `\&12\catcode `\#12\catcode `\^12\catcode `\_12\catcode `\%12\relax}%
\providecommand \@@startlink[1]{}%
\providecommand \@@endlink[0]{}%
\providecommand \url  [0]{\begingroup\@sanitize@url \@url }%
\providecommand \@url [1]{\endgroup\@href {#1}{\urlprefix }}%
\providecommand \urlprefix  [0]{URL }%
\providecommand \Eprint [0]{\href }%
\providecommand \doibase [0]{http://dx.doi.org/}%
\providecommand \selectlanguage [0]{\@gobble}%
\providecommand \bibinfo  [0]{\@secondoftwo}%
\providecommand \bibfield  [0]{\@secondoftwo}%
\providecommand \translation [1]{[#1]}%
\providecommand \BibitemOpen [0]{}%
\providecommand \bibitemStop [0]{}%
\providecommand \bibitemNoStop [0]{.\EOS\space}%
\providecommand \EOS [0]{\spacefactor3000\relax}%
\providecommand \BibitemShut  [1]{\csname bibitem#1\endcsname}%
\let\auto@bib@innerbib\@empty
\bibitem [{\citenamefont {Abbott}\ \emph {et~al.}(2017)\citenamefont {Abbott},
  \citenamefont {Abbott}, \citenamefont {Abbott}, \citenamefont {Acernese},
  \citenamefont {Ackley}, \citenamefont {Adams}, \citenamefont {Adams},
  \citenamefont {Addesso}, \citenamefont {Adhikari}, \citenamefont {Adya},\
  and\ \citenamefont {et~al.}}]{Abbott_2017}%
  \BibitemOpen
  \bibfield  {author} {\bibinfo {author} {\bibfnamefont {B.}~\bibnamefont
  {Abbott}}, \bibinfo {author} {\bibfnamefont {R.}~\bibnamefont {Abbott}},
  \bibinfo {author} {\bibfnamefont {T.}~\bibnamefont {Abbott}}, \bibinfo
  {author} {\bibfnamefont {F.}~\bibnamefont {Acernese}}, \bibinfo {author}
  {\bibfnamefont {K.}~\bibnamefont {Ackley}}, \bibinfo {author} {\bibfnamefont
  {C.}~\bibnamefont {Adams}}, \bibinfo {author} {\bibfnamefont
  {T.}~\bibnamefont {Adams}}, \bibinfo {author} {\bibfnamefont
  {P.}~\bibnamefont {Addesso}}, \bibinfo {author} {\bibfnamefont
  {R.}~\bibnamefont {Adhikari}}, \bibinfo {author} {\bibfnamefont
  {V.}~\bibnamefont {Adya}}, \ and\ \bibinfo {author} {\bibnamefont {et~al.}},\
  }\href {\doibase 10.1103/physrevlett.119.161101} {\bibfield  {journal}
  {\bibinfo  {journal} {Physical Review Letters}\ }\textbf {\bibinfo {volume}
  {119}} (\bibinfo {year} {2017}),\ 10.1103/physrevlett.119.161101}\BibitemShut
  {NoStop}%
\bibitem [{\citenamefont {Abbott}\ \emph {et~al.}(2016)\citenamefont {Abbott}
  \emph {et~al.}}]{Abbott:2016blz}%
  \BibitemOpen
  \bibfield  {author} {\bibinfo {author} {\bibfnamefont {B.~P.}\ \bibnamefont
  {Abbott}} \emph {et~al.} (\bibinfo {collaboration} {LIGO Scientific,
  Virgo}),\ }\href {\doibase 10.1103/PhysRevLett.116.061102} {\bibfield
  {journal} {\bibinfo  {journal} {Phys. Rev. Lett.}\ }\textbf {\bibinfo
  {volume} {116}},\ \bibinfo {pages} {061102} (\bibinfo {year} {2016})},\
  \Eprint {http://arxiv.org/abs/1602.03837} {arXiv:1602.03837 [gr-qc]}
  \BibitemShut {NoStop}%
\bibitem [{\citenamefont {Lindblom}\ \emph {et~al.}(2008)\citenamefont
  {Lindblom}, \citenamefont {Owen},\ and\ \citenamefont
  {Brown}}]{Lindblom_2008}%
  \BibitemOpen
  \bibfield  {author} {\bibinfo {author} {\bibfnamefont {L.}~\bibnamefont
  {Lindblom}}, \bibinfo {author} {\bibfnamefont {B.~J.}\ \bibnamefont {Owen}},
  \ and\ \bibinfo {author} {\bibfnamefont {D.~A.}\ \bibnamefont {Brown}},\
  }\href {\doibase 10.1103/physrevd.78.124020} {\bibfield  {journal} {\bibinfo
  {journal} {Physical Review D}\ }\textbf {\bibinfo {volume} {78}} (\bibinfo
  {year} {2008}),\ 10.1103/physrevd.78.124020}\BibitemShut {NoStop}%
\bibitem [{\citenamefont {Aasi}\ \emph {et~al.}(2015)\citenamefont {Aasi},
  \citenamefont {Abbott}, \citenamefont {Abbott}, \citenamefont {Abbott},
  \citenamefont {Abernathy}, \citenamefont {Ackley}, \citenamefont {Adams},
  \citenamefont {Adams}, \citenamefont {Addesso},\ and\ \citenamefont
  {et~al.}}]{2015}%
  \BibitemOpen
  \bibfield  {author} {\bibinfo {author} {\bibfnamefont {J.}~\bibnamefont
  {Aasi}}, \bibinfo {author} {\bibfnamefont {B.~P.}\ \bibnamefont {Abbott}},
  \bibinfo {author} {\bibfnamefont {R.}~\bibnamefont {Abbott}}, \bibinfo
  {author} {\bibfnamefont {T.}~\bibnamefont {Abbott}}, \bibinfo {author}
  {\bibfnamefont {M.~R.}\ \bibnamefont {Abernathy}}, \bibinfo {author}
  {\bibfnamefont {K.}~\bibnamefont {Ackley}}, \bibinfo {author} {\bibfnamefont
  {C.}~\bibnamefont {Adams}}, \bibinfo {author} {\bibfnamefont
  {T.}~\bibnamefont {Adams}}, \bibinfo {author} {\bibfnamefont
  {P.}~\bibnamefont {Addesso}}, \ and\ \bibinfo {author} {\bibnamefont
  {et~al.}},\ }\href {\doibase 10.1088/0264-9381/32/7/074001} {\bibfield
  {journal} {\bibinfo  {journal} {Classical and Quantum Gravity}\ }\textbf
  {\bibinfo {volume} {32}},\ \bibinfo {pages} {074001} (\bibinfo {year}
  {2015})}\BibitemShut {NoStop}%
\bibitem [{\citenamefont {Acernese}\ \emph {et~al.}(2014)\citenamefont
  {Acernese}, \citenamefont {Agathos}, \citenamefont {Agatsuma}, \citenamefont
  {Aisa}, \citenamefont {Allemandou}, \citenamefont {Allocca}, \citenamefont
  {Amarni}, \citenamefont {Astone}, \citenamefont {Balestri}, \citenamefont
  {Ballardin},\ and\ \citenamefont {et~al.}}]{Acernese_2014}%
  \BibitemOpen
  \bibfield  {author} {\bibinfo {author} {\bibfnamefont {F.}~\bibnamefont
  {Acernese}}, \bibinfo {author} {\bibfnamefont {M.}~\bibnamefont {Agathos}},
  \bibinfo {author} {\bibfnamefont {K.}~\bibnamefont {Agatsuma}}, \bibinfo
  {author} {\bibfnamefont {D.}~\bibnamefont {Aisa}}, \bibinfo {author}
  {\bibfnamefont {N.}~\bibnamefont {Allemandou}}, \bibinfo {author}
  {\bibfnamefont {A.}~\bibnamefont {Allocca}}, \bibinfo {author} {\bibfnamefont
  {J.}~\bibnamefont {Amarni}}, \bibinfo {author} {\bibfnamefont
  {P.}~\bibnamefont {Astone}}, \bibinfo {author} {\bibfnamefont
  {G.}~\bibnamefont {Balestri}}, \bibinfo {author} {\bibfnamefont
  {G.}~\bibnamefont {Ballardin}}, \ and\ \bibinfo {author} {\bibnamefont
  {et~al.}},\ }\href {\doibase 10.1088/0264-9381/32/2/024001} {\bibfield
  {journal} {\bibinfo  {journal} {Classical and Quantum Gravity}\ }\textbf
  {\bibinfo {volume} {32}},\ \bibinfo {pages} {024001} (\bibinfo {year}
  {2014})}\BibitemShut {NoStop}%
\bibitem [{\citenamefont {Amaro-Seoane}\ \emph {et~al.}(2017)\citenamefont
  {Amaro-Seoane}, \citenamefont {Audley}, \citenamefont {Babak}, \citenamefont
  {Baker}, \citenamefont {Barausse}, \citenamefont {Bender}, \citenamefont
  {Berti}, \citenamefont {Binetruy}, \citenamefont {Born},\ and\ \citenamefont
  {et~al.}}]{amaroseoane2017laser}%
  \BibitemOpen
  \bibfield  {author} {\bibinfo {author} {\bibfnamefont {P.}~\bibnamefont
  {Amaro-Seoane}}, \bibinfo {author} {\bibfnamefont {H.}~\bibnamefont
  {Audley}}, \bibinfo {author} {\bibfnamefont {S.}~\bibnamefont {Babak}},
  \bibinfo {author} {\bibfnamefont {J.}~\bibnamefont {Baker}}, \bibinfo
  {author} {\bibfnamefont {E.}~\bibnamefont {Barausse}}, \bibinfo {author}
  {\bibfnamefont {P.}~\bibnamefont {Bender}}, \bibinfo {author} {\bibfnamefont
  {E.}~\bibnamefont {Berti}}, \bibinfo {author} {\bibfnamefont
  {P.}~\bibnamefont {Binetruy}}, \bibinfo {author} {\bibfnamefont
  {M.}~\bibnamefont {Born}}, \ and\ \bibinfo {author} {\bibfnamefont {D.~B.}\
  \bibnamefont {et~al.}},\ }\href@noop {} {\enquote {\bibinfo {title} {Laser
  interferometer space antenna},}\ } (\bibinfo {year} {2017}),\ \Eprint
  {http://arxiv.org/abs/1702.00786} {arXiv:1702.00786 [astro-ph.IM]}
  \BibitemShut {NoStop}%
\bibitem [{\citenamefont {Sathyaprakash}\ \emph {et~al.}(2012)\citenamefont
  {Sathyaprakash}, \citenamefont {Abernathy}, \citenamefont {Acernese},
  \citenamefont {Ajith}, \citenamefont {Allen}, \citenamefont {Amaro-Seoane},
  \citenamefont {Andersson}, \citenamefont {Aoudia}, \citenamefont {Arun},
  \citenamefont {Astone},\ and\ \citenamefont {et~al.}}]{Sathyaprakash_2012}%
  \BibitemOpen
  \bibfield  {author} {\bibinfo {author} {\bibfnamefont {B.}~\bibnamefont
  {Sathyaprakash}}, \bibinfo {author} {\bibfnamefont {M.}~\bibnamefont
  {Abernathy}}, \bibinfo {author} {\bibfnamefont {F.}~\bibnamefont {Acernese}},
  \bibinfo {author} {\bibfnamefont {P.}~\bibnamefont {Ajith}}, \bibinfo
  {author} {\bibfnamefont {B.}~\bibnamefont {Allen}}, \bibinfo {author}
  {\bibfnamefont {P.}~\bibnamefont {Amaro-Seoane}}, \bibinfo {author}
  {\bibfnamefont {N.}~\bibnamefont {Andersson}}, \bibinfo {author}
  {\bibfnamefont {S.}~\bibnamefont {Aoudia}}, \bibinfo {author} {\bibfnamefont
  {K.}~\bibnamefont {Arun}}, \bibinfo {author} {\bibfnamefont {P.}~\bibnamefont
  {Astone}}, \ and\ \bibinfo {author} {\bibnamefont {et~al.}},\ }\href
  {\doibase 10.1088/0264-9381/29/12/124013} {\bibfield  {journal} {\bibinfo
  {journal} {Classical and Quantum Gravity}\ }\textbf {\bibinfo {volume}
  {29}},\ \bibinfo {pages} {124013} (\bibinfo {year} {2012})}\BibitemShut
  {NoStop}%
\bibitem [{\citenamefont {Blanchet}(2014)}]{Blanchet:2013haa}%
  \BibitemOpen
  \bibfield  {author} {\bibinfo {author} {\bibfnamefont {L.}~\bibnamefont
  {Blanchet}},\ }\href {\doibase 10.12942/lrr-2014-2} {\bibfield  {journal}
  {\bibinfo  {journal} {Living Rev. Rel.}\ }\textbf {\bibinfo {volume} {17}},\
  \bibinfo {pages} {2} (\bibinfo {year} {2014})},\ \Eprint
  {http://arxiv.org/abs/1310.1528} {arXiv:1310.1528 [gr-qc]} \BibitemShut
  {NoStop}%
\bibitem [{\citenamefont {{Grandcl{\'e}ment}}\ and\ \citenamefont
  {{Novak}}(2009)}]{2009LRR....12....1G}%
  \BibitemOpen
  \bibfield  {author} {\bibinfo {author} {\bibfnamefont {P.}~\bibnamefont
  {{Grandcl{\'e}ment}}}\ and\ \bibinfo {author} {\bibfnamefont
  {J.}~\bibnamefont {{Novak}}},\ }\href {\doibase 10.12942/lrr-2009-1}
  {\bibfield  {journal} {\bibinfo  {journal} {Living Reviews in Relativity}\
  }\textbf {\bibinfo {volume} {12}},\ \bibinfo {eid} {1} (\bibinfo {year}
  {2009})},\ \Eprint {http://arxiv.org/abs/0706.2286} {arXiv:0706.2286 [gr-qc]}
  \BibitemShut {NoStop}%
\bibitem [{\citenamefont {Goldberger}\ and\ \citenamefont
  {Rothstein}(2006)}]{goldberger_effective_2006}%
  \BibitemOpen
  \bibfield  {author} {\bibinfo {author} {\bibfnamefont {W.~D.}\ \bibnamefont
  {Goldberger}}\ and\ \bibinfo {author} {\bibfnamefont {I.~Z.}\ \bibnamefont
  {Rothstein}},\ }\href {\doibase 10.1103/PhysRevD.73.104029} {\bibfield
  {journal} {\bibinfo  {journal} {Physical Review D}\ }\textbf {\bibinfo
  {volume} {73}} (\bibinfo {year} {2006}),\ 10.1103/PhysRevD.73.104029},\
  \bibinfo {note} {arXiv: hep-th/0409156}\BibitemShut {NoStop}%
\bibitem [{\citenamefont {Porto}(2016)}]{porto_effective_2016}%
  \BibitemOpen
  \bibfield  {author} {\bibinfo {author} {\bibfnamefont {R.~A.}\ \bibnamefont
  {Porto}},\ }\href {\doibase 10.1016/j.physrep.2016.04.003} {\bibfield
  {journal} {\bibinfo  {journal} {Physics Reports}\ }\textbf {\bibinfo {volume}
  {633}},\ \bibinfo {pages} {1} (\bibinfo {year} {2016})},\ \bibinfo {note}
  {arXiv: 1601.04914}\BibitemShut {NoStop}%
\bibitem [{\citenamefont {Levi}(2018)}]{Levi:2018nxp}%
  \BibitemOpen
  \bibfield  {author} {\bibinfo {author} {\bibfnamefont {M.}~\bibnamefont
  {Levi}},\ }\href@noop {} {\  (\bibinfo {year} {2018})},\ \Eprint
  {http://arxiv.org/abs/1807.01699} {arXiv:1807.01699 [hep-th]} \BibitemShut
  {NoStop}%
\bibitem [{\citenamefont {Foffa}\ and\ \citenamefont
  {Sturani}(2013)}]{Foffa_2013}%
  \BibitemOpen
  \bibfield  {author} {\bibinfo {author} {\bibfnamefont {S.}~\bibnamefont
  {Foffa}}\ and\ \bibinfo {author} {\bibfnamefont {R.}~\bibnamefont
  {Sturani}},\ }\href {\doibase 10.1103/physrevd.87.064011} {\bibfield
  {journal} {\bibinfo  {journal} {Physical Review D}\ }\textbf {\bibinfo
  {volume} {87}} (\bibinfo {year} {2013}),\
  10.1103/physrevd.87.064011}\BibitemShut {NoStop}%
\bibitem [{\citenamefont {Foffa}\ \emph {et~al.}(2017)\citenamefont {Foffa},
  \citenamefont {Mastrolia}, \citenamefont {Sturani},\ and\ \citenamefont
  {Sturm}}]{Foffa_2017}%
  \BibitemOpen
  \bibfield  {author} {\bibinfo {author} {\bibfnamefont {S.}~\bibnamefont
  {Foffa}}, \bibinfo {author} {\bibfnamefont {P.}~\bibnamefont {Mastrolia}},
  \bibinfo {author} {\bibfnamefont {R.}~\bibnamefont {Sturani}}, \ and\
  \bibinfo {author} {\bibfnamefont {C.}~\bibnamefont {Sturm}},\ }\href
  {\doibase 10.1103/physrevd.95.104009} {\bibfield  {journal} {\bibinfo
  {journal} {Physical Review D}\ }\textbf {\bibinfo {volume} {95}} (\bibinfo
  {year} {2017}),\ 10.1103/physrevd.95.104009}\BibitemShut {NoStop}%
\bibitem [{\citenamefont {{Foffa}}\ and\ \citenamefont
  {{Sturani}}(2019)}]{2019PhRvD.100b4047F}%
  \BibitemOpen
  \bibfield  {author} {\bibinfo {author} {\bibfnamefont {S.}~\bibnamefont
  {{Foffa}}}\ and\ \bibinfo {author} {\bibfnamefont {R.}~\bibnamefont
  {{Sturani}}},\ }\href {\doibase 10.1103/PhysRevD.100.024047} {\bibfield
  {journal} {\bibinfo  {journal} {\prd}\ }\textbf {\bibinfo {volume} {100}},\
  \bibinfo {eid} {024047} (\bibinfo {year} {2019})},\ \Eprint
  {http://arxiv.org/abs/1903.05113} {arXiv:1903.05113 [gr-qc]} \BibitemShut
  {NoStop}%
\bibitem [{\citenamefont {{Foffa}}\ \emph {et~al.}(2019)\citenamefont
  {{Foffa}}, \citenamefont {{Porto}}, \citenamefont {{Rothstein}},\ and\
  \citenamefont {{Sturani}}}]{2019PhRvD.100b4048F}%
  \BibitemOpen
  \bibfield  {author} {\bibinfo {author} {\bibfnamefont {S.}~\bibnamefont
  {{Foffa}}}, \bibinfo {author} {\bibfnamefont {R.~A.}\ \bibnamefont
  {{Porto}}}, \bibinfo {author} {\bibfnamefont {I.}~\bibnamefont
  {{Rothstein}}}, \ and\ \bibinfo {author} {\bibfnamefont {R.}~\bibnamefont
  {{Sturani}}},\ }\href {\doibase 10.1103/PhysRevD.100.024048} {\bibfield
  {journal} {\bibinfo  {journal} {\prd}\ }\textbf {\bibinfo {volume} {100}},\
  \bibinfo {eid} {024048} (\bibinfo {year} {2019})},\ \Eprint
  {http://arxiv.org/abs/1903.05118} {arXiv:1903.05118 [gr-qc]} \BibitemShut
  {NoStop}%
\bibitem [{\citenamefont {Damour}\ \emph {et~al.}(2014)\citenamefont {Damour},
  \citenamefont {Jaranowski},\ and\ \citenamefont {Sch{\"a}fer}}]{Damour_2014}%
  \BibitemOpen
  \bibfield  {author} {\bibinfo {author} {\bibfnamefont {T.}~\bibnamefont
  {Damour}}, \bibinfo {author} {\bibfnamefont {P.}~\bibnamefont {Jaranowski}},
  \ and\ \bibinfo {author} {\bibfnamefont {G.}~\bibnamefont {Sch{\"a}fer}},\
  }\href {\doibase 10.1103/physrevd.89.064058} {\bibfield  {journal} {\bibinfo
  {journal} {Physical Review D}\ }\textbf {\bibinfo {volume} {89}} (\bibinfo
  {year} {2014}),\ 10.1103/physrevd.89.064058}\BibitemShut {NoStop}%
\bibitem [{\citenamefont {Damour}\ \emph {et~al.}(2015)\citenamefont {Damour},
  \citenamefont {Jaranowski},\ and\ \citenamefont {Sch{\"a}fer}}]{Damour_2015}%
  \BibitemOpen
  \bibfield  {author} {\bibinfo {author} {\bibfnamefont {T.}~\bibnamefont
  {Damour}}, \bibinfo {author} {\bibfnamefont {P.}~\bibnamefont {Jaranowski}},
  \ and\ \bibinfo {author} {\bibfnamefont {G.}~\bibnamefont {Sch{\"a}fer}},\
  }\href {\doibase 10.1103/physrevd.91.084024} {\bibfield  {journal} {\bibinfo
  {journal} {Physical Review D}\ }\textbf {\bibinfo {volume} {91}} (\bibinfo
  {year} {2015}),\ 10.1103/physrevd.91.084024}\BibitemShut {NoStop}%
\bibitem [{\citenamefont {Damour}\ \emph {et~al.}(2016)\citenamefont {Damour},
  \citenamefont {Jaranowski},\ and\ \citenamefont {Sch{\"a}fer}}]{Damour_2016}%
  \BibitemOpen
  \bibfield  {author} {\bibinfo {author} {\bibfnamefont {T.}~\bibnamefont
  {Damour}}, \bibinfo {author} {\bibfnamefont {P.}~\bibnamefont {Jaranowski}},
  \ and\ \bibinfo {author} {\bibfnamefont {G.}~\bibnamefont {Sch{\"a}fer}},\
  }\href {\doibase 10.1103/physrevd.93.084014} {\bibfield  {journal} {\bibinfo
  {journal} {Physical Review D}\ }\textbf {\bibinfo {volume} {93}} (\bibinfo
  {year} {2016}),\ 10.1103/physrevd.93.084014}\BibitemShut {NoStop}%
\bibitem [{\citenamefont {Bernard}\ \emph {et~al.}(2016)\citenamefont
  {Bernard}, \citenamefont {Blanchet}, \citenamefont {Boh{\'e}}, \citenamefont
  {Faye},\ and\ \citenamefont {Marsat}}]{Bernard_2016}%
  \BibitemOpen
  \bibfield  {author} {\bibinfo {author} {\bibfnamefont {L.}~\bibnamefont
  {Bernard}}, \bibinfo {author} {\bibfnamefont {L.}~\bibnamefont {Blanchet}},
  \bibinfo {author} {\bibfnamefont {A.}~\bibnamefont {Boh{\'e}}}, \bibinfo
  {author} {\bibfnamefont {G.}~\bibnamefont {Faye}}, \ and\ \bibinfo {author}
  {\bibfnamefont {S.}~\bibnamefont {Marsat}},\ }\href {\doibase
  10.1103/physrevd.93.084037} {\bibfield  {journal} {\bibinfo  {journal}
  {Physical Review D}\ }\textbf {\bibinfo {volume} {93}} (\bibinfo {year}
  {2016}),\ 10.1103/physrevd.93.084037}\BibitemShut {NoStop}%
\bibitem [{\citenamefont {Bernard}\ \emph
  {et~al.}(2017{\natexlab{a}})\citenamefont {Bernard}, \citenamefont
  {Blanchet}, \citenamefont {Boh{\'e}}, \citenamefont {Faye},\ and\
  \citenamefont {Marsat}}]{Bernard_2017}%
  \BibitemOpen
  \bibfield  {author} {\bibinfo {author} {\bibfnamefont {L.}~\bibnamefont
  {Bernard}}, \bibinfo {author} {\bibfnamefont {L.}~\bibnamefont {Blanchet}},
  \bibinfo {author} {\bibfnamefont {A.}~\bibnamefont {Boh{\'e}}}, \bibinfo
  {author} {\bibfnamefont {G.}~\bibnamefont {Faye}}, \ and\ \bibinfo {author}
  {\bibfnamefont {S.}~\bibnamefont {Marsat}},\ }\href {\doibase
  10.1103/physrevd.95.044026} {\bibfield  {journal} {\bibinfo  {journal}
  {Physical Review D}\ }\textbf {\bibinfo {volume} {95}} (\bibinfo {year}
  {2017}{\natexlab{a}}),\ 10.1103/physrevd.95.044026}\BibitemShut {NoStop}%
\bibitem [{\citenamefont {Bernard}\ \emph
  {et~al.}(2017{\natexlab{b}})\citenamefont {Bernard}, \citenamefont
  {Blanchet}, \citenamefont {Boh{\'e}}, \citenamefont {Faye},\ and\
  \citenamefont {Marsat}}]{Bernard_2017_2}%
  \BibitemOpen
  \bibfield  {author} {\bibinfo {author} {\bibfnamefont {L.}~\bibnamefont
  {Bernard}}, \bibinfo {author} {\bibfnamefont {L.}~\bibnamefont {Blanchet}},
  \bibinfo {author} {\bibfnamefont {A.}~\bibnamefont {Boh{\'e}}}, \bibinfo
  {author} {\bibfnamefont {G.}~\bibnamefont {Faye}}, \ and\ \bibinfo {author}
  {\bibfnamefont {S.}~\bibnamefont {Marsat}},\ }\href {\doibase
  10.1103/physrevd.96.104043} {\bibfield  {journal} {\bibinfo  {journal}
  {Physical Review D}\ }\textbf {\bibinfo {volume} {96}} (\bibinfo {year}
  {2017}{\natexlab{b}}),\ 10.1103/physrevd.96.104043}\BibitemShut {NoStop}%
\bibitem [{\citenamefont {Marchand}\ \emph {et~al.}(2018)\citenamefont
  {Marchand}, \citenamefont {Bernard}, \citenamefont {Blanchet},\ and\
  \citenamefont {Faye}}]{Marchand_2018}%
  \BibitemOpen
  \bibfield  {author} {\bibinfo {author} {\bibfnamefont {T.}~\bibnamefont
  {Marchand}}, \bibinfo {author} {\bibfnamefont {L.}~\bibnamefont {Bernard}},
  \bibinfo {author} {\bibfnamefont {L.}~\bibnamefont {Blanchet}}, \ and\
  \bibinfo {author} {\bibfnamefont {G.}~\bibnamefont {Faye}},\ }\href {\doibase
  10.1103/physrevd.97.044023} {\bibfield  {journal} {\bibinfo  {journal}
  {Physical Review D}\ }\textbf {\bibinfo {volume} {97}} (\bibinfo {year}
  {2018}),\ 10.1103/physrevd.97.044023}\BibitemShut {NoStop}%
\bibitem [{\citenamefont {Bernard}\ \emph {et~al.}(2018)\citenamefont
  {Bernard}, \citenamefont {Blanchet}, \citenamefont {Faye},\ and\
  \citenamefont {Marchand}}]{Bernard_2018}%
  \BibitemOpen
  \bibfield  {author} {\bibinfo {author} {\bibfnamefont {L.}~\bibnamefont
  {Bernard}}, \bibinfo {author} {\bibfnamefont {L.}~\bibnamefont {Blanchet}},
  \bibinfo {author} {\bibfnamefont {G.}~\bibnamefont {Faye}}, \ and\ \bibinfo
  {author} {\bibfnamefont {T.}~\bibnamefont {Marchand}},\ }\href {\doibase
  10.1103/physrevd.97.044037} {\bibfield  {journal} {\bibinfo  {journal}
  {Physical Review D}\ }\textbf {\bibinfo {volume} {97}} (\bibinfo {year}
  {2018}),\ 10.1103/physrevd.97.044037}\BibitemShut {NoStop}%
\bibitem [{\citenamefont {Porto}(2006)}]{Porto_2006}%
  \BibitemOpen
  \bibfield  {author} {\bibinfo {author} {\bibfnamefont {R.~A.}\ \bibnamefont
  {Porto}},\ }\href {\doibase 10.1103/physrevd.73.104031} {\bibfield  {journal}
  {\bibinfo  {journal} {Physical Review D}\ }\textbf {\bibinfo {volume} {73}}
  (\bibinfo {year} {2006}),\ 10.1103/physrevd.73.104031}\BibitemShut {NoStop}%
\bibitem [{\citenamefont {Levi}\ and\ \citenamefont
  {Steinhoff}(2015{\natexlab{a}})}]{Levi:2015msa}%
  \BibitemOpen
  \bibfield  {author} {\bibinfo {author} {\bibfnamefont {M.}~\bibnamefont
  {Levi}}\ and\ \bibinfo {author} {\bibfnamefont {J.}~\bibnamefont
  {Steinhoff}},\ }\href {\doibase 10.1007/JHEP09(2015)219} {\bibfield
  {journal} {\bibinfo  {journal} {JHEP}\ }\textbf {\bibinfo {volume} {09}},\
  \bibinfo {pages} {219} (\bibinfo {year} {2015}{\natexlab{a}})},\ \Eprint
  {http://arxiv.org/abs/1501.04956} {arXiv:1501.04956 [gr-qc]} \BibitemShut
  {NoStop}%
\bibitem [{\citenamefont {Levi}\ \emph {et~al.}(2020)\citenamefont {Levi},
  \citenamefont {McLeod},\ and\ \citenamefont {von Hippel}}]{Levi:2020kvb}%
  \BibitemOpen
  \bibfield  {author} {\bibinfo {author} {\bibfnamefont {M.}~\bibnamefont
  {Levi}}, \bibinfo {author} {\bibfnamefont {A.~J.}\ \bibnamefont {McLeod}}, \
  and\ \bibinfo {author} {\bibfnamefont {M.}~\bibnamefont {von Hippel}},\
  }\href@noop {} {\  (\bibinfo {year} {2020})},\ \Eprint
  {http://arxiv.org/abs/2003.02827} {arXiv:2003.02827 [hep-th]} \BibitemShut
  {NoStop}%
\bibitem [{\citenamefont {Levi}\ and\ \citenamefont
  {Steinhoff}(2015{\natexlab{b}})}]{Levi:2014gsa}%
  \BibitemOpen
  \bibfield  {author} {\bibinfo {author} {\bibfnamefont {M.}~\bibnamefont
  {Levi}}\ and\ \bibinfo {author} {\bibfnamefont {J.}~\bibnamefont
  {Steinhoff}},\ }\href {\doibase 10.1007/JHEP06(2015)059} {\bibfield
  {journal} {\bibinfo  {journal} {JHEP}\ }\textbf {\bibinfo {volume} {06}},\
  \bibinfo {pages} {059} (\bibinfo {year} {2015}{\natexlab{b}})},\ \Eprint
  {http://arxiv.org/abs/1410.2601} {arXiv:1410.2601 [gr-qc]} \BibitemShut
  {NoStop}%
\bibitem [{\citenamefont {Levi}\ and\ \citenamefont
  {Steinhoff}(2016{\natexlab{a}})}]{Levi:2015ixa}%
  \BibitemOpen
  \bibfield  {author} {\bibinfo {author} {\bibfnamefont {M.}~\bibnamefont
  {Levi}}\ and\ \bibinfo {author} {\bibfnamefont {J.}~\bibnamefont
  {Steinhoff}},\ }\href {\doibase 10.1088/1475-7516/2016/01/008} {\bibfield
  {journal} {\bibinfo  {journal} {JCAP}\ }\textbf {\bibinfo {volume} {1601}},\
  \bibinfo {pages} {008} (\bibinfo {year} {2016}{\natexlab{a}})},\ \Eprint
  {http://arxiv.org/abs/1506.05794} {arXiv:1506.05794 [gr-qc]} \BibitemShut
  {NoStop}%
\bibitem [{\citenamefont {Levi}\ and\ \citenamefont
  {Steinhoff}(2016{\natexlab{b}})}]{Levi:2016ofk}%
  \BibitemOpen
  \bibfield  {author} {\bibinfo {author} {\bibfnamefont {M.}~\bibnamefont
  {Levi}}\ and\ \bibinfo {author} {\bibfnamefont {J.}~\bibnamefont
  {Steinhoff}},\ }\href@noop {} {\  (\bibinfo {year} {2016}{\natexlab{b}})},\
  \Eprint {http://arxiv.org/abs/1607.04252} {arXiv:1607.04252 [gr-qc]}
  \BibitemShut {NoStop}%
\bibitem [{\citenamefont {{Cheung}}(2017)}]{2017arXiv170803872C}%
  \BibitemOpen
  \bibfield  {author} {\bibinfo {author} {\bibfnamefont {C.}~\bibnamefont
  {{Cheung}}},\ }\href@noop {} {\bibfield  {journal} {\bibinfo  {journal}
  {arXiv e-prints}\ ,\ \bibinfo {eid} {arXiv:1708.03872}} (\bibinfo {year}
  {2017})},\ \Eprint {http://arxiv.org/abs/1708.03872} {arXiv:1708.03872
  [hep-ph]} \BibitemShut {NoStop}%
\bibitem [{\citenamefont {Bern}\ \emph {et~al.}(2019)\citenamefont {Bern},
  \citenamefont {Cheung}, \citenamefont {Roiban}, \citenamefont {Shen},
  \citenamefont {Solon},\ and\ \citenamefont {Zeng}}]{PhysRevLett.122.201603}%
  \BibitemOpen
  \bibfield  {author} {\bibinfo {author} {\bibfnamefont {Z.}~\bibnamefont
  {Bern}}, \bibinfo {author} {\bibfnamefont {C.}~\bibnamefont {Cheung}},
  \bibinfo {author} {\bibfnamefont {R.}~\bibnamefont {Roiban}}, \bibinfo
  {author} {\bibfnamefont {C.-H.}\ \bibnamefont {Shen}}, \bibinfo {author}
  {\bibfnamefont {M.~P.}\ \bibnamefont {Solon}}, \ and\ \bibinfo {author}
  {\bibfnamefont {M.}~\bibnamefont {Zeng}},\ }\href {\doibase
  10.1103/PhysRevLett.122.201603} {\bibfield  {journal} {\bibinfo  {journal}
  {Phys. Rev. Lett.}\ }\textbf {\bibinfo {volume} {122}},\ \bibinfo {pages}
  {201603} (\bibinfo {year} {2019})}\BibitemShut {NoStop}%
\bibitem [{\citenamefont {Cheung}\ \emph {et~al.}(2018)\citenamefont {Cheung},
  \citenamefont {Rothstein},\ and\ \citenamefont {Solon}}]{Cheung:2018wkq}%
  \BibitemOpen
  \bibfield  {author} {\bibinfo {author} {\bibfnamefont {C.}~\bibnamefont
  {Cheung}}, \bibinfo {author} {\bibfnamefont {I.~Z.}\ \bibnamefont
  {Rothstein}}, \ and\ \bibinfo {author} {\bibfnamefont {M.~P.}\ \bibnamefont
  {Solon}},\ }\href {\doibase 10.1103/PhysRevLett.121.251101} {\bibfield
  {journal} {\bibinfo  {journal} {Phys. Rev. Lett.}\ }\textbf {\bibinfo
  {volume} {121}},\ \bibinfo {pages} {251101} (\bibinfo {year} {2018})},\
  \Eprint {http://arxiv.org/abs/1808.02489} {arXiv:1808.02489 [hep-th]}
  \BibitemShut {NoStop}%
\bibitem [{\citenamefont {Antonelli}\ \emph {et~al.}(2019)\citenamefont
  {Antonelli}, \citenamefont {Buonanno}, \citenamefont {Steinhoff},
  \citenamefont {van~de Meent},\ and\ \citenamefont
  {Vines}}]{Antonelli:2019ytb}%
  \BibitemOpen
  \bibfield  {author} {\bibinfo {author} {\bibfnamefont {A.}~\bibnamefont
  {Antonelli}}, \bibinfo {author} {\bibfnamefont {A.}~\bibnamefont {Buonanno}},
  \bibinfo {author} {\bibfnamefont {J.}~\bibnamefont {Steinhoff}}, \bibinfo
  {author} {\bibfnamefont {M.}~\bibnamefont {van~de Meent}}, \ and\ \bibinfo
  {author} {\bibfnamefont {J.}~\bibnamefont {Vines}},\ }\href {\doibase
  10.1103/PhysRevD.99.104004} {\bibfield  {journal} {\bibinfo  {journal} {Phys.
  Rev.}\ }\textbf {\bibinfo {volume} {D99}},\ \bibinfo {pages} {104004}
  (\bibinfo {year} {2019})},\ \Eprint {http://arxiv.org/abs/1901.07102}
  {arXiv:1901.07102 [gr-qc]} \BibitemShut {NoStop}%
\bibitem [{\citenamefont {Polyakov}(1981)}]{Polyakov}%
  \BibitemOpen
  \bibfield  {author} {\bibinfo {author} {\bibfnamefont {A.}~\bibnamefont
  {Polyakov}},\ }\href {\doibase 10.1016/0370-2693(81)90743-7} {\bibfield
  {journal} {\bibinfo  {journal} {Physics Letters B}\ }\textbf {\bibinfo
  {volume} {103}},\ \bibinfo {pages} {207} (\bibinfo {year}
  {1981})}\BibitemShut {NoStop}%
\bibitem [{\citenamefont {Galley}\ and\ \citenamefont
  {Porto}(2013)}]{Galley_2013}%
  \BibitemOpen
  \bibfield  {author} {\bibinfo {author} {\bibfnamefont {C.~R.}\ \bibnamefont
  {Galley}}\ and\ \bibinfo {author} {\bibfnamefont {R.~A.}\ \bibnamefont
  {Porto}},\ }\href {\doibase 10.1007/jhep11(2013)096} {\bibfield  {journal}
  {\bibinfo  {journal} {Journal of High Energy Physics}\ }\textbf {\bibinfo
  {volume} {2013}} (\bibinfo {year} {2013}),\
  10.1007/jhep11(2013)096}\BibitemShut {NoStop}%
\bibitem [{\citenamefont {{Davis}}\ and\ \citenamefont
  {{Melville}}(2019)}]{2019arXiv191008831D}%
  \BibitemOpen
  \bibfield  {author} {\bibinfo {author} {\bibfnamefont {A.-C.}\ \bibnamefont
  {{Davis}}}\ and\ \bibinfo {author} {\bibfnamefont {S.}~\bibnamefont
  {{Melville}}},\ }\href@noop {} {\bibfield  {journal} {\bibinfo  {journal}
  {arXiv e-prints}\ ,\ \bibinfo {eid} {arXiv:1910.08831}} (\bibinfo {year}
  {2019})},\ \Eprint {http://arxiv.org/abs/1910.08831} {arXiv:1910.08831
  [gr-qc]} \BibitemShut {NoStop}%
\bibitem [{\citenamefont {Detweiler}(2008)}]{Detweiler:2008ft}%
  \BibitemOpen
  \bibfield  {author} {\bibinfo {author} {\bibfnamefont {S.~L.}\ \bibnamefont
  {Detweiler}},\ }\href {\doibase 10.1103/PhysRevD.77.124026} {\bibfield
  {journal} {\bibinfo  {journal} {Phys. Rev.}\ }\textbf {\bibinfo {volume}
  {D77}},\ \bibinfo {pages} {124026} (\bibinfo {year} {2008})},\ \Eprint
  {http://arxiv.org/abs/0804.3529} {arXiv:0804.3529 [gr-qc]} \BibitemShut
  {NoStop}%
\bibitem [{Note1()}]{Note1}%
  \BibitemOpen
  \bibinfo {note} {Defined as the smallest possible circular orbit ; not to be
  confused with the Innermost \protect \textit {stable} Circular Orbit or
  ISCO}\BibitemShut {NoStop}%
\bibitem [{\citenamefont {Foffa}(2014)}]{Foffa:2013gja}%
  \BibitemOpen
  \bibfield  {author} {\bibinfo {author} {\bibfnamefont {S.}~\bibnamefont
  {Foffa}},\ }\href {\doibase 10.1103/PhysRevD.89.024019} {\bibfield  {journal}
  {\bibinfo  {journal} {Phys. Rev.}\ }\textbf {\bibinfo {volume} {D89}},\
  \bibinfo {pages} {024019} (\bibinfo {year} {2014})},\ \Eprint
  {http://arxiv.org/abs/1309.3956} {arXiv:1309.3956 [gr-qc]} \BibitemShut
  {NoStop}%
\bibitem [{\citenamefont {Wheeler}(1945)}]{Wheeler:1945aa}%
  \BibitemOpen
  \bibfield  {author} {\bibinfo {author} {\bibfnamefont {J.~A.}\ \bibnamefont
  {Wheeler}},\ }\href {\doibase 10.1103/RevModPhys.17.157} {\bibfield
  {journal} {\bibinfo  {journal} {Reviews of Modern Physics}\ }\textbf
  {\bibinfo {volume} {17}},\ \bibinfo {pages} {157} (\bibinfo {year}
  {1945})}\BibitemShut {NoStop}%
\bibitem [{\citenamefont {Gilmore}\ and\ \citenamefont
  {Ross}(2008)}]{Gilmore_2008}%
  \BibitemOpen
  \bibfield  {author} {\bibinfo {author} {\bibfnamefont {J.~B.}\ \bibnamefont
  {Gilmore}}\ and\ \bibinfo {author} {\bibfnamefont {A.}~\bibnamefont {Ross}},\
  }\href {\doibase 10.1103/physrevd.78.124021} {\bibfield  {journal} {\bibinfo
  {journal} {Physical Review D}\ }\textbf {\bibinfo {volume} {78}} (\bibinfo
  {year} {2008}),\ 10.1103/physrevd.78.124021}\BibitemShut {NoStop}%
\bibitem [{\citenamefont {Kol}\ and\ \citenamefont
  {Smolkin}(2008)}]{Kol:2007rx}%
  \BibitemOpen
  \bibfield  {author} {\bibinfo {author} {\bibfnamefont {B.}~\bibnamefont
  {Kol}}\ and\ \bibinfo {author} {\bibfnamefont {M.}~\bibnamefont {Smolkin}},\
  }\href {\doibase 10.1103/PhysRevD.77.064033} {\bibfield  {journal} {\bibinfo
  {journal} {Phys.Rev.D}\ }\textbf {\bibinfo {volume} {77}},\ \bibinfo {pages}
  {064033} (\bibinfo {year} {2008})},\ \Eprint {http://arxiv.org/abs/0712.2822}
  {arXiv:0712.2822 [hep-th]} \BibitemShut {NoStop}%
\bibitem [{\citenamefont {Blanchet}\ \emph {et~al.}(2011)\citenamefont
  {Blanchet}, \citenamefont {Detweiler}, \citenamefont {Le~Tiec},\ and\
  \citenamefont {Whiting}}]{Blanchet:2011aha}%
  \BibitemOpen
  \bibfield  {author} {\bibinfo {author} {\bibfnamefont {L.}~\bibnamefont
  {Blanchet}}, \bibinfo {author} {\bibfnamefont {S.}~\bibnamefont {Detweiler}},
  \bibinfo {author} {\bibfnamefont {A.}~\bibnamefont {Le~Tiec}}, \ and\
  \bibinfo {author} {\bibfnamefont {B.~F.}\ \bibnamefont {Whiting}},\
  }\bibfield  {booktitle} {\emph {\bibinfo {booktitle} {{Mass and motion in
  general relativity. Proceedings, School on Mass, Orleans, France, June 23-25,
  2008}}},\ }\href {\doibase 10.1007/978-90-481-3015-3_15} {\bibfield
  {journal} {\bibinfo  {journal} {Fundam. Theor. Phys.}\ }\textbf {\bibinfo
  {volume} {162}},\ \bibinfo {pages} {415} (\bibinfo {year} {2011})},\ \bibinfo
  {note} {[,415(2010)]},\ \Eprint {http://arxiv.org/abs/1007.2614}
  {arXiv:1007.2614 [gr-qc]} \BibitemShut {NoStop}%
\bibitem [{\citenamefont {K{\"a}lin}\ and\ \citenamefont
  {Porto}(2020)}]{Kalin:2019rwq}%
  \BibitemOpen
  \bibfield  {author} {\bibinfo {author} {\bibfnamefont {G.}~\bibnamefont
  {K{\"a}lin}}\ and\ \bibinfo {author} {\bibfnamefont {R.~A.}\ \bibnamefont
  {Porto}},\ }\href {\doibase 10.1007/JHEP01(2020)072} {\bibfield  {journal}
  {\bibinfo  {journal} {JHEP}\ }\textbf {\bibinfo {volume} {01}},\ \bibinfo
  {pages} {072} (\bibinfo {year} {2020})},\ \Eprint
  {http://arxiv.org/abs/1910.03008} {arXiv:1910.03008 [hep-th]} \BibitemShut
  {NoStop}%
\bibitem [{Note2()}]{Note2}%
  \BibitemOpen
  \bibinfo {note} {Due to the symmetry of our equations, $e_1$ and $e_2$ become
  complex-valued at the same value of the frequency $\omega $}\BibitemShut
  {NoStop}%
\bibitem [{\citenamefont {Buonanno}\ and\ \citenamefont
  {Damour}(1999)}]{Buonanno:1998gg}%
  \BibitemOpen
  \bibfield  {author} {\bibinfo {author} {\bibfnamefont {A.}~\bibnamefont
  {Buonanno}}\ and\ \bibinfo {author} {\bibfnamefont {T.}~\bibnamefont
  {Damour}},\ }\href {\doibase 10.1103/PhysRevD.59.084006} {\bibfield
  {journal} {\bibinfo  {journal} {Phys.Rev.D}\ }\textbf {\bibinfo {volume}
  {59}},\ \bibinfo {pages} {084006} (\bibinfo {year} {1999})},\ \Eprint
  {http://arxiv.org/abs/gr-qc/9811091} {arXiv:gr-qc/9811091} \BibitemShut
  {NoStop}%
\end{thebibliography}%

\end{document}